\documentclass[aps,prb,reprint,twocolumn,superscriptaddress,floatfix]{revtex4-2}
\usepackage{amsmath}
\usepackage{amssymb}
\usepackage[utf8]{inputenc}
\usepackage{makeidx}
\usepackage{graphicx}
\usepackage[caption=false]{subfig}
\usepackage{hyperref} 
\usepackage{siunitx}
\usepackage{xcolor}
\usepackage{braket}
\usepackage{booktabs}
\usepackage{makecell}
\usepackage[normalem]{ulem}
\usepackage{newfloat}
\usepackage{dcolumn}
\usepackage{mathtools}
\usepackage{natbib}

\makeatletter
\newcommand*{\balancecolsandclearpage}{%
  \close@column@grid
  \clearpage
  \twocolumngrid
}
\makeatother

\pdfoutput=1

\begin{document}

\title{Origin of Nonlinear Damping due to Mode Coupling in Auto-Oscillatory Modes Strongly Driven by Spin-Orbit Torque}

\author{Inhee Lee}
\email{lee.2338@osu.edu}
\affiliation{Department of Physics, The Ohio State University, Columbus, OH 43210, USA}

\author{Chi Zhang}
\affiliation{Department of Physics, The Ohio State University, Columbus, OH 43210, USA}

\author{Simranjeet Singh}
 \altaffiliation[Present address: ]{Department of Physics, Carnegie Mellon university, Pittsburgh,
 PA, 15213, USA}
\affiliation{Department of Physics, The Ohio State University,
Columbus, OH 43210, USA}

\author{Brendan McCullian}
\affiliation{Department of Physics, The Ohio State University, Columbus, OH 43210, USA}

\author{P. Chris Hammel}
\email{hammel@physics.osu.edu}
\affiliation{Department of Physics, The Ohio State University, Columbus, OH 43210, USA}

\date{\today}

\begin{abstract}
We investigate the physical origin of nonlinear damping due to
mode coupling between several auto-oscillatory modes driven by
spin-orbit torque in constricted Py/Pt heterostructures by
examining the dependence of auto-oscillation on temperature and
applied field orientation. We observe a transition in the
nonlinear damping of the auto-oscillation modes extracted from the
total oscillation power as a function of drive current, which
coincides with the onset of power redistribution amongst several
modes and the crossover from linewidth narrowing to linewidth
broadening in all individual modes. This indicates the activation
of another relaxation process by nonlinear magnon-magnon
scattering within the modes. We also find that both nonlinear
damping and threshold current in the mode-interaction damping
regime at high drive current after transition are temperature
independent, suggesting that the mode coupling occurs dominantly
through a non-thermal magnon scattering process via a dipole or
exchange interaction rather than thermally excited magnon-mediated
scattering. This finding presents a promising pathway to overcome
the current limitations of efficiently controlling the interaction
between two highly nonlinear magnetic oscillators to prevent mode
crosstalk or inter-mode energy transfer and deepens understanding
of complex nonlinear spin dynamics in multimode spin wave systems.
\end{abstract}

\maketitle

Spin-orbit torque driven magnetic nano-oscillators have recently
emerged as potential charge current tunable microwave sources for
spintronics
devices\cite{Demidov2012,Liu2013,Demidov2014,Duan2014,Yang2015,
Collet2016,Mazraati2016,Awad2017,Dvornik2018,Mazraati2018,Fulara2019,Haidar2019,Safranski2019},
as well as fundamental elements for neuromorphic computing
\cite{Zahedinejad2020}. These oscillators use spin Hall effect to
convert the charge current into a pure spin current which is
injected into the ferromagnet, exerting an anti-damping torque on
the magnetization. Above the threshold current, coherent magnetic
auto-oscillations (AO) are generated at microwave frequencies.

In principle, these planar spin Hall nano-oscillators (SHNO) need
not be limited in size and are expected to provide larger AO power
because the Oersted field is relatively small and uniform in the
device unlike nano-pillar spin-torque nano-oscillators (STNO).
However, additional damping channels arise in extended thin film
structures through nonlinear magnon scattering that prevents
auto-oscillation \cite{Duan2014}. These damping channels can be
suppressed by restricting the area of the AO by fabricating a
nano-constriction \cite{Demidov2014, Mazraati2018} as well as
other methods of spatial mode confinement such as dimensional
reduction \cite{Duan2014, Verba2020} and local dipolar field
\cite{Lee2010,Lee2011,Zhang2019,Zhang2021}.

Nevertheless, in constriction-based SHNOs, mode splittings are
often observed \cite{Demidov2014, Mazraati2018}, leading to
substantial linewidth broadening which degrades performance in
terms of AO coherence and power. So far, little is known about the
origins of nonlinear damping as a consequence of multimode
excitation in this system, and the mechanism underlying mode
coupling is still unknown. In nano-pillar STNOs, similar multimode
behaviors such as mode hopping \cite{Bonetti2010, Muduli2012PRL,
Muduli2012PRB, Bonetti2012, Heinonen2013, Iacocca2014, Zhang2016},
mode coexistence \cite{Sankey2005,Dumas2013,Iacocca2015} and
3-magnon process \cite{Barsukov2019,Etesamirad2021}, have been
reported along with theoretical studies
\cite{deAguiar2007,Bonetti2010,Muduli2012PRL,Muduli2012PRB,Bonetti2012,Heinonen2013,Iacocca2014,Iacocca2015,Zhang2016,Zhang2017}
of mode coupling, including its description \cite{Muduli2012PRL,
Muduli2012PRB, Bonetti2012, Heinonen2013, Iacocca2014, Zhang2016,
Zhang2017}, its effect on mode decoherence
\cite{Muduli2012PRB,Heinonen2013,Iacocca2014} and its temperature
dependence
\cite{Muduli2012PRB,Heinonen2013,Iacocca2014,Zhang2017}, however,
much remains to be understood.

\begin{figure}
\includegraphics[width=1.0\columnwidth]{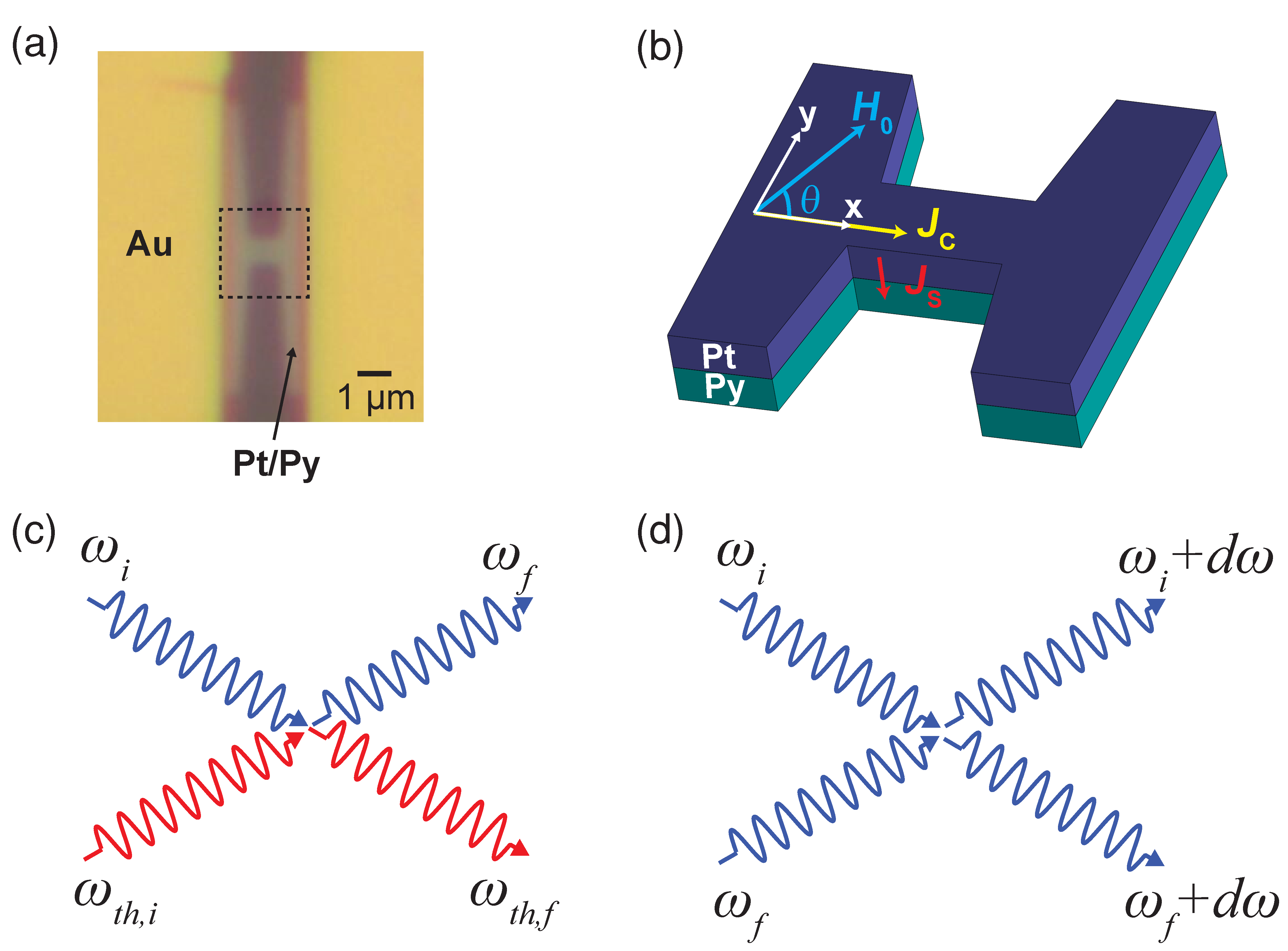}
\caption{(a) Optical image of the auto-oscillation device. (b)
Schematic diagram of the 2D constricted Pt(5 nm)/Py(5 nm) bilayer
structure of the bow tie shape in the dashed box of (a). Due to
the strong spin-orbit coupling in Pt, the charge current density
$\mathbf{J_{c}}$ flowing in the x direction along the axis of
microstrip line is converted to pure spin current density
$\mathbf{J_{s}}$ which is injected into the ferromagnet Py with
the appropriate spin polarization required for anti-damping
torque. $\theta$ is the angle made by the direction of in-plane
applied field $\mathbf{H_{0}}$ relative to the x-axis. (c) Four
magnon scattering mediated by thermally excited magnons
$\omega_{th,i}$ and $\omega_{th,f}$ in equilibrium with thermal
reservoir. $\omega_{i} + \omega_{th,i} = \omega_{f} +
\omega_{th,f}$. (d) Four magnon scattering by intermode
interactions. $d \omega$ is the frequency shift within the mode
caused by dipole or exchange interactions between $\omega_{i}$ and
$\omega_{j}$.} \label{fig:sample structure}
\end{figure}

For this study, we fabricate Pt(5nm)/Py(5nm) bilayer devices in
the form of a bow tie with an active center area of 600 nm
$\times$ 1 $\mu$m, as shown in Fig. \ref{fig:sample structure}(a).
This sample structure allows the generation of multiple AO modes
by applying a high current density $J_c$ that is converted into a
spin current $J_s$ that, in turn, exerts a spin torque on the
magnetization of the Py (See Fig. \ref{fig:sample structure}(b)).
By changing the orientation of the magnetic field $\mathbf{H_{0}}$
applied at angle $\theta$ with respect to the x-axis of the
microstrip line, we tune the eigenmodes of the spin waves defined
by the spatial pattern of the internal field in Py, which is
significantly modified by its dipolar, or demagnetizing field.

\begin{figure*}
\includegraphics[width=2.0\columnwidth]{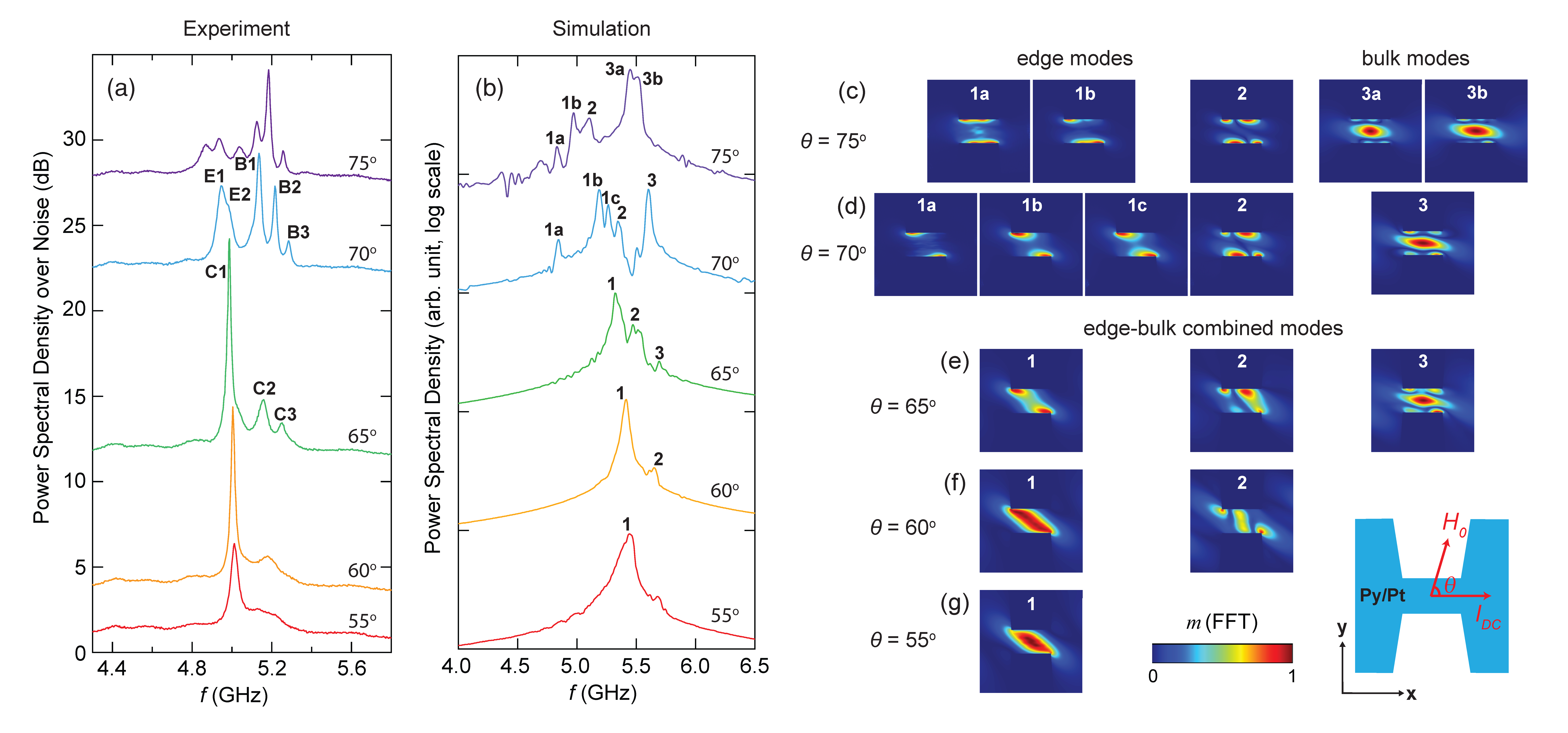}
\caption{(a) Angle dependence of the auto-oscillation spectrum
measured at $I_{\rm DC} = 7$ mA, $T = 77$ K and $H_{0} = 570$ Oe.
We vary the angle $\theta$ between $\bf{H_0}$ and $I_{\rm DC}$ in
our experiments as shown in the diagram in the lower right corner
of the figure. In the spectra for $65^{\circ}$ and $70^{\circ}$,
`C', `E' and `B' represent a combined edge-bulk mode, edge mode,
and bulk mode, respectively. (b) Spectra for various $\theta$
obtained from micromagnetic simulations using Mumax3 performed
with the current density distribution and Oersted field in Fig.
\ref{fig:J_HOe_gaus} of Appendix. The corresponding $I_{\rm DC}$
are 7.92 mA for $\theta=55^{\circ}$, 7.26 mA for
$\theta=60^{\circ}$, 6.92 mA for $\theta=65^{\circ}$, 6.52 mA for
$\theta=70^{\circ}$, and 6.36 mA for $\theta=75^{\circ}$, about
0.2 - 0.3 mA larger than the respective threshold current at each
$\theta$. In (a) and (b), each spectrum is offset. (c)-(g) Spatial
eigenmode profiles corresponding to the spectral peaks indicated
by the numbers in (b) at various $\theta$. The dynamic
magnetization amplitude $m$ is normalized on the color scale of
each image. At $\theta=70^{\circ}$ and $75^{\circ}$, the spatial
separation of edge (1-2) modes and bulk (3) modes occurs as shown
in (c) and (d) as well as their spectral separation as shown in
(b).} \label{fig:spectrum}
\end{figure*}

Indeed, when the orientation of the applied field changes relative
to the device edges causing mode constriction, spectral and
spatial distributions of the resulting AO modes vary
significantly. Fig. \ref{fig:spectrum}(a) shows the angle $\theta$
dependence of the AO spectrum measured with bias current $I_{\rm
DC} =$ 7 mA at temperature $T = 77$ K in an applied field $H_0 =
570$ Oe. As $\theta$ increases, more excitation modes appear over
a wider frequency range as a result of stronger mode constriction.
We perform micromagnetic simulations using MuMax3
\cite{Mumax3}(see Appendix \ref{section:simulation}) to understand
the complex spectra at various $\theta$ and to identify the
relevant AO modes. Fig. \ref{fig:spectrum}(b) shows the angle
dependence of the AO spectra obtained from simulations conducted
with the current density distribution and Oersted field in Fig. 10
of Appendix. It well describes the overall evolution of the
experimental AO spectrum with increasing angle in Fig.
\ref{fig:spectrum}(a), although more accurate spatial information
reflecting the high current density nonlinearity and structural
defects in the AO active region seems necessary to fully account
for the spectral shape details. Fig. \ref{fig:spectrum}(c)-(g)
show the spatial profiles of the spin wave eigenmodes
corresponding to the spectral peaks in Fig. \ref{fig:spectrum}(b)
at various angles $\theta$. We find that for $\theta \leq
65^{\circ}$, the edge and bulk modes are combined, whereas for
$\theta \geq 70^{\circ}$, the edge and bulk modes are spatially
separated due to the significant difference in the demagnetizing
field between the edge and center regions. This separation also
occurs spectrally (see Fig. \ref{fig:spectrum}(a) and
\ref{fig:spectrum}(b)) and the largest AO peak appears in one of
the bulk modes at frequencies higher than edge modes, which is in
good agreement with previous AO results obtained on nanowires with
$\theta = 85^{\circ}$ \cite{Duan2014}.

The spectral and spatial mode profiles for AO in the active mode
shown in Fig. \ref{fig:spectrum} differ significantly from those
of the linear spin wave mode (see Fig.
\ref{fig:linear_spin_wave_mode} in Appendix). This demonstrates
that the dynamics of the AO mode are different from those of the
linear spin wave mode, arising from nonlinear effects such as
complex bullet mode dynamics involved with mode size
reduction\cite{Demidov2012,Demidov2014,Slavin2005} or mode size
oscillation\cite{Yang2015,Wagner2018}. In particular, this bullet
mode effect appears to be more pronounced in bulk mode 3a and 3b
in Fig. \ref{fig:spectrum}(c) and bulk mode 3 in Fig.
\ref{fig:spectrum}(d), away from the edge effect. The frequency
jump as the signature of the bullet mode is also discussed in
Appendix \ref{section:STFMR}.

\begin{figure}
\includegraphics[width=1.0\columnwidth]{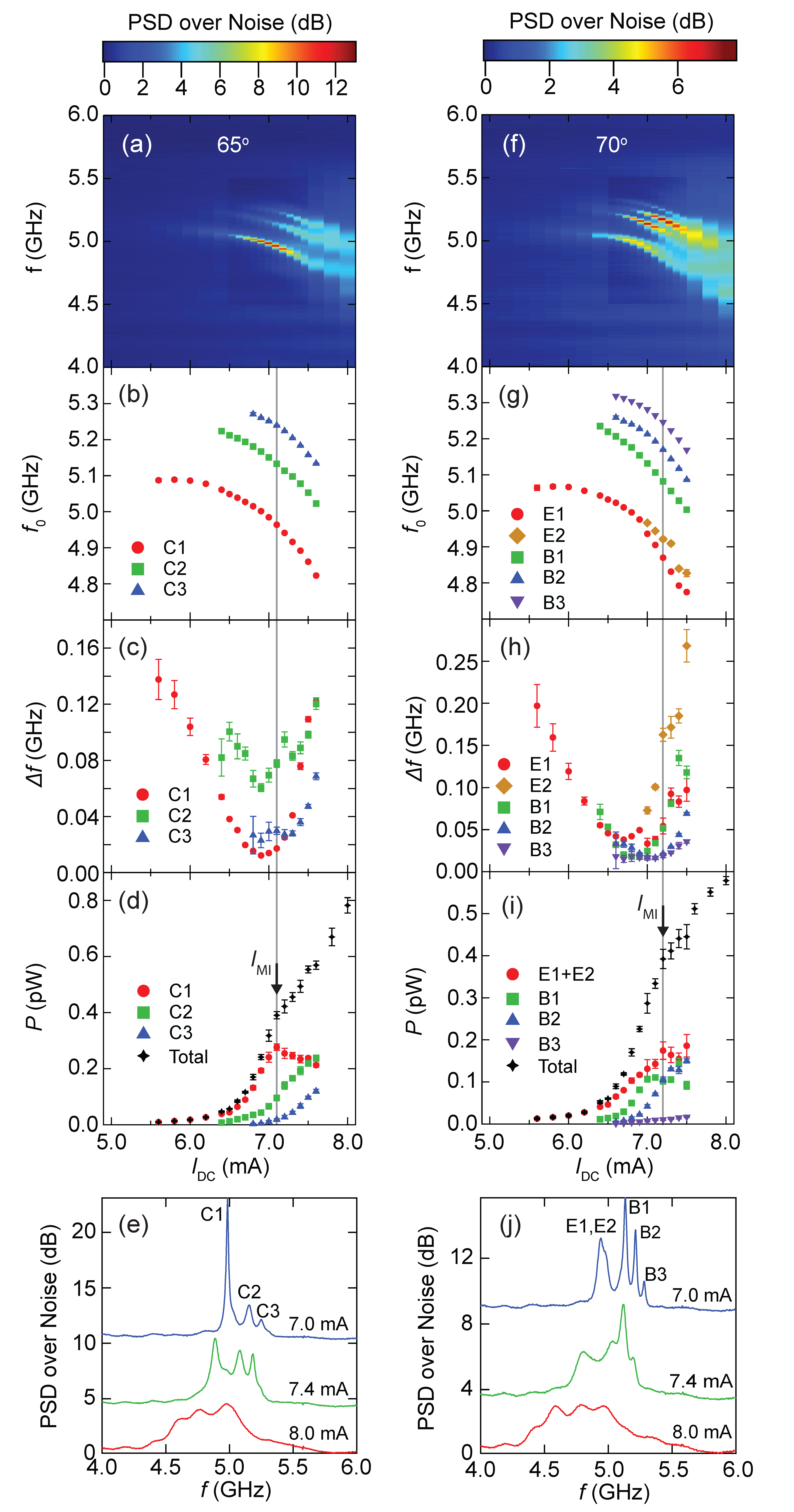}
\caption{(a) Evolution of power spectral density (PSD) with
increasing bias current $I_{\rm DC}$, (b) resonance frequency
$f_{0}$ vs. $I_{\rm DC}$, (c) linewidth $\Delta f$ vs. $I_{\rm
DC}$, (d) power $P$ vs. $I_{\rm DC}$ for the main modes C1, C2 and
C3 for $\theta = 65^{\circ}$. The FMR parameters in (b)-(d) are
extracted by fitting the data in (a) to Lorentzian functions. (e)
Representative spectra at high $I_{\rm DC}$ for $\theta =
65^{\circ}$ showing the redistribution of power among modes at
high $I_{\rm DC}$. (f) Evolution of PSD with increasing $I_{\rm
DC}$, (g) $f_{0}$ vs. $I_{\rm DC}$, (h) $\Delta f$ vs. $I_{\rm
DC}$, (i) $P$ vs. $I_{\rm DC}$ of the main edge (E1, E2) and bulk
(B1, B2, B3) modes for $\theta = 70^{\circ}$. The FMR parameters
in (g)-(i) are extracted by fitting the data in (f) to Lorentzian
functions. (j) Representative spectra at high $I_{\rm DC}$ for
$\theta = 70^{\circ}$ showing the redistribution of power among
modes at high drive current. In (d) and (i), the black diamond
markers are the total power $P_{\rm total}$ summed over all
excited modes, and $I_{\rm MI}$ represents the current at which
the nonlinear damping transition occurs. At very high $I_{\rm
DC}$, the main modes shown here are not clearly identified due to
the large linewidth broadening and the emergence of other excited
modes even though $P_{\rm total}$ can be obtained. Here $H_{0} =
570$ Oe and $T = 77$ K for all measurements.} \label{fig:fmr
parameters}
\end{figure}

In the evolution of the auto-oscillation modes with increasing
bias current, we observe a transition in nonlinear damping due to
mode-mode coupling. As can be seen in Fig. \ref{fig:fmr
parameters}(a) obtained by measuring at $\theta = 65^{\circ}$, as
the current $I_{\rm DC}$--and hence the anti-damping torque--
increases. the three main modes C1, C2, and C3 (labelled in Fig.
\ref{fig:fmr parameters}(e)) appear in the spectrum above each
threshold current. We characterize these modes evolving at various
$I_{\rm DC}$ with the resonance frequency $f_0$, linewidth $\Delta
f$, and power $P$ obtained from Lorentzian fits, which are shown
in Fig. \ref{fig:fmr parameters}(b)-(d). Associated with the
transition of nonlinear damping, we observe noticeable abrupt
changes at the current $I_{\rm MI} = 7.1$ mA, such as the turnover
from the linewidth narrowing to linewidth broadening (sign change
of $\Delta f / I_{\rm DC}$) for all excitation modes (Fig.
\ref{fig:fmr parameters}(c)) and power saturation in C1 mode (Fig.
\ref{fig:fmr parameters}(d)). This transition marks the onset of
another relaxation process with additional damping. However, as
$I_{\rm DC}$ increases in the high current regime ($I_{\rm DC}
> I_{\rm MI}$), we also observe i) a monotonic redshift of the frequency for all three modes reflecting a monotonic reduction in saturation magnetization due to magnon excitation \cite{Demidov2011} (Fig. \ref{fig:fmr parameters}(b)),
ii) a power increase in the C2 and C3 modes despite their
linewidth broadening (Fig. \ref{fig:fmr parameters}(c) and
\ref{fig:fmr parameters}(d)), and iii) an increase in the total
power (Fig. \ref{fig:fmr parameters} (d)). All of these imply that
the magnon population is growing faster than excited magnons can
decay to other thermal reservoirs. Therefore, we conclude that
nonlinear damping in the high current regime occurs through magnon
redistribution from low-frequency modes to high-frequency modes
via nonlinear magnon-magnon scattering as shown schematically in
Fig. \ref{fig:sample structure}(c) or \ref{fig:sample
structure}(d).

There are other possible causes for the nonlinear damping
transition at $I_{\rm MI}$. The transition in the dependence of
power on current shown in Fig. \ref{fig:fmr parameters}(d) may be
described by the selective excitation and selective saturation of
each mode. However, C2 and C3 in multimode excitation cannot be
explained this way because their linewidth broadening, along with
the power increase, at $I_{\rm DC} \geq I_{\rm MI}$ is not
consistent with typical single-mode AO behavior which exhibits
linewidth narrowing with increasing $I_{\rm DC}$ as C1 does for
$I_{\rm DC} < I_{\rm MI}$. Power saturation of C1 may be caused by
Joule heating or spin pumping. However, Joule heating cannot
saturate the power of C1 without similarly affecting C2 and C3.
Crucially, Joule heating can be observed in the spin torque FMR by
applying a high negative current ($I_{\rm DC} < 0$) corresponding
to the positive damping in which the spin current driven magnons
are not generated. As the absolute value of $I_{\rm DC} < 0$
increases, the decreasing resonance field can change to an
increasing resonance field due to the decrease in saturation
magnetization caused by Joule heating, similar to the previous
Brillouin light scattering results \cite{Demidov2011}. However,
this effect does not appear up to $I_{\rm DC}$ = -8 mA in our data
(see Fig. \ref{fig:IDC Dep of STFMR} in Appendix), showing that
Joule heating has a small effect. The intensity of spin pumping is
proportional to the power of the AO mode \cite{Jungfleisch2015}.
As shown in Fig. \ref{fig:fmr parameters}(c) and \ref{fig:fmr
parameters}(d), the linewidths of C1, C2 and C3 with different
powers start to increase together at $I_{\rm MI}$ with increasing
$I_{\rm DC}$, which cannot be explained by spin pumping.
Therefore, we conclude that only mode-mode interaction can explain
the coincidence of both the power transfer from C1 to the higher
frequency modes C2 and C3 and the turnover from linewidth
broadening to linewidth narrowing of all three modes at $I_{\rm
MI}$ for increasing $I_{\rm DC}$, eventually leading to an almost
even power distribution in the spectrum at 8 mA, as shown in Fig.
\ref{fig:fmr parameters}(e).

\begin{figure}
\includegraphics[width=1.0\columnwidth]{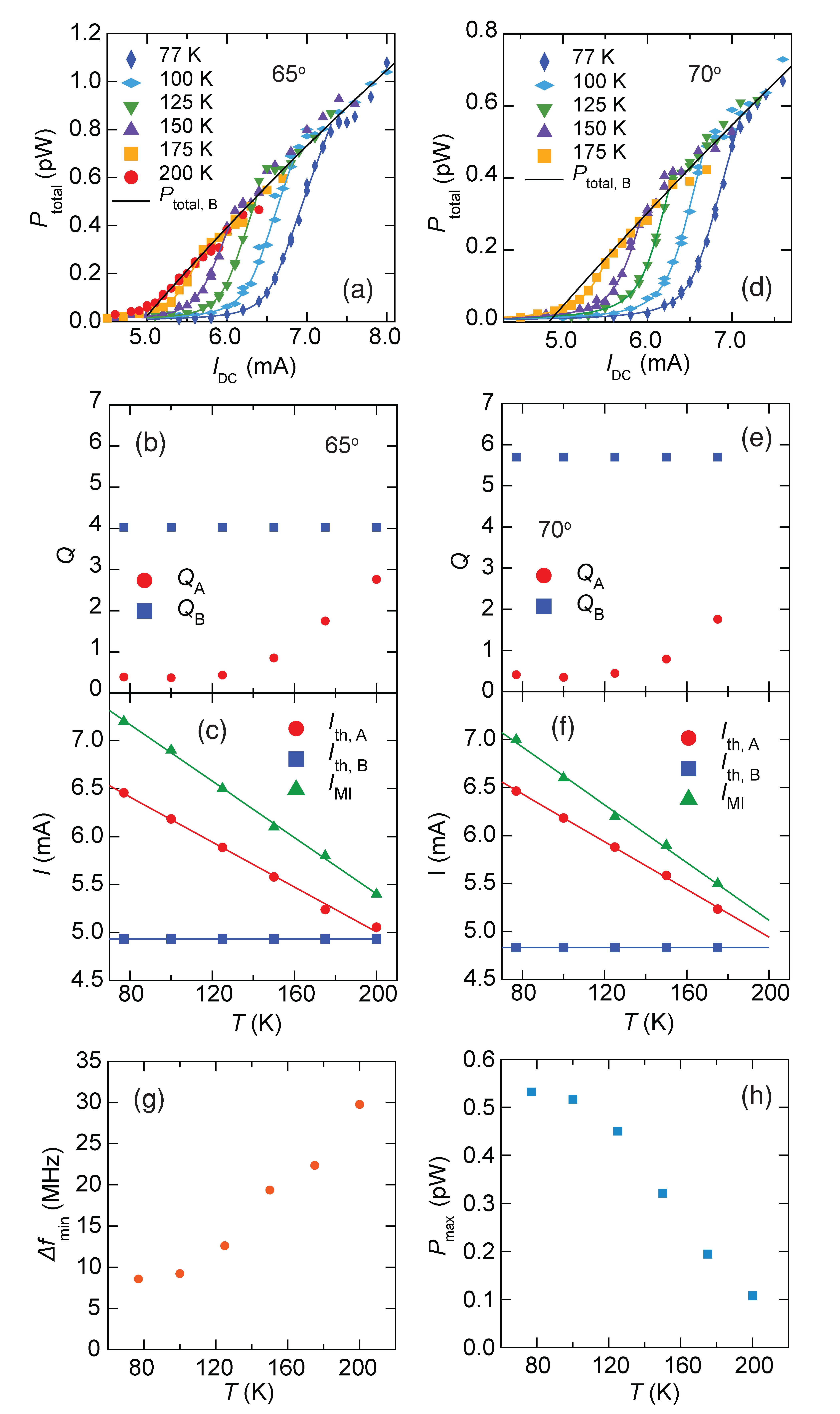}
\caption{(a) Total power $P_{\rm total}$ of the AO as a function
of $I_{\rm DC}$ at various temperature $T$, (b) the nonlinear
damping coefficient $Q$ vs. $T$, (c) the threshold current $I_{\rm
th}$ vs. $T$ and the transition current $I_{\rm MI}$ vs. $T$ for
$\theta = 65^{\circ}$. Colored solid curves are the linear fits to
Eq. (\ref{eqn:Ith}) with $I_{\rm th,0} = 7.35 \; \rm{mA}$ and
$\kappa_{\rm th} = -0.0117 \; \rm{mA/K}$ and Eq. (\ref{eqn:IMI})
with $I_{\rm MI,0} = 8.34 \; \rm{mA}$ and $\kappa_{\rm MI} =
-0.0147 \; \rm{mA/K}$ for $I_{\rm th}$ and $I_{\rm MI}$,
respectively. Similarly, (d) $P_{\rm total}$ of the AO as a
function of $I_{\rm DC}$ at various $T$, (e) $Q$ vs. $T$, (f)
$I_{\rm th}$ vs. $T$ and $I_{\rm MI}$ vs. $T$ for $\theta =
70^{\circ}$. Colored solid curves are the linear fits to Eq.
(\ref{eqn:Ith}) with $I_{\rm th,0} = 7.43 \; \rm{mA}$ and
$\kappa_{\rm th} = -0.0124 \; \rm{mA/K}$ and Eq. (\ref{eqn:IMI})
with $I_{\rm MI,0} = 8.12 \; \rm{mA}$ and $\kappa_{\rm MI} =
-0.015 \; \rm{mA/K}$ for $I_{\rm th}$ and $I_{\rm MI}$,
respectively. Colored solid curves in Fig. \ref{fig:nonlinear
parameters}(a) and \ref{fig:nonlinear parameters}(d) are the
$P_{\rm total}$ calculated using the theoretical equation (see Eq.
(\ref{eqn:PtotallIMI}) in Appendix \ref{section:Q}) with $Q_A$ and
$I_{\rm th,A}$ in \ref{fig:nonlinear parameters}(b)-(c) and
\ref{fig:nonlinear parameters}(e)-(f), respectively. The black
curves in Fig. \ref{fig:nonlinear parameters}(a) and
\ref{fig:nonlinear parameters}(d) are the $P_{\rm total}$
calculated using the theoretical equation (see Eq.
(\ref{eqn:PtotalgeqIMI}) in Appendix \ref{section:Q}) with the
common values of $Q_B$ and $I_{\rm th,B}$ in \ref{fig:nonlinear
parameters}(b)-(c), \ref{fig:nonlinear parameters}(e)-(f),
respectively. Here `A' represents individual mode dominant regime
($I_{\rm DC} < I_{\rm MI}$) and `B' represents mode interaction
activation regime ($I_{\rm DC} \geq I_{\rm MI}$). And the
normalization factor $N_{0} =$ 9.5 pW and 8.7 pW, determined from
fitting, for $\theta = 65^{\circ}$ and $\theta = 70^{\circ}$,
respectively, (see Appendix \ref{section:Q}) and $H_{\rm 0} = 490$
Oe. Temperature dependence of (g) minimum linewidth and (h)
maximum power of C1 mode for $\theta = 65^{\circ}$.}
\label{fig:nonlinear parameters}
\end{figure}

In Fig. \ref{fig:nonlinear parameters} we show the temperature
dependence of the nonlinear damping of the AO system along with
the threshold current $I_{\rm th}$ and transition current $I_{\rm
MI}$. Fig. \ref{fig:nonlinear parameters}(a) shows the total power
$P_{\rm total}$ summed for all excitation modes as a function of
$I_{\rm DC}$ obtained at various temperatures for $\theta =
65^{\circ}$. For each temperature, $P_{\rm total}$ has two current
regimes, `A' for $I_{\rm DC} < I_{\rm MI}$ in which a single mode
is dominant, and `B' for $I_{\rm DC} \geq I_{\rm MI}$ in which
mode-mode interaction is active amongst multiple excitation modes,
separated by the kink in $P_{\rm total}$ at which $P$ in the
lowest frequency mode saturates. One of our remarkable findings is
that total power $P_{\rm total, B}$ for all temperatures in regime
B falls onto a single common curve (black line in Fig.
\ref{fig:nonlinear parameters}(a)), whereas total power $P_{\rm
total, A}$ in regime A is temperature dependent. This suggests
that the mode-mode coupling occurs through a non-thermal process
(Fig. \ref{fig:sample structure}(d)) originating from exchange or
dipole interactions instead of the thermally excited
magnon-mediated scattering (Fig. \ref{fig:sample structure}(c)).
The rapid growth of $P_{\rm total,A}$ (color lines in Fig.
\ref{fig:nonlinear parameters}(a)) from each threshold current
$I_{\rm th}$ is eventually limited by $P_{\rm total,B}$ (black
line in Fig. \ref{fig:nonlinear parameters}(a)) at each
corresponding transition current $I_{\rm MI}$ for all
temperatures. This indicates that the much faster relaxation
process arising from the temperature-independent mode-mode
interactions in regime B predominates over the temperature
dependent single mode relaxation process occurring in regime A.
Note that magnon thermalization through redistribution within the
dynamic magnetic system is possible only if the relaxation via
nonlinear magnon scattering arising from mode coupling is much
faster than the relaxation to the external non-magnetic systems in
individual modes by intrinsic Gilbert damping or spin pumping.

In order to quantify the nonlinear damping of the AO system, we
discuss the parameter $Q$ that represents the change in positive
nonlinear damping $\Gamma_{+} \approx \Gamma_{\rm G} \left( 1 + Q
p \right)$ with increasing AO power, where $\Gamma_{\rm G}$ is the
Gilbert damping \cite{Slavin2009}. $Q$ as well as $I_{\rm th}$ can
be obtained from the relationship between the normalized AO power
$p$ and $I_{\rm DC}$ (see Appendix \ref{section:Q}). Fig.
\ref{fig:nonlinear parameters}(b) and \ref{fig:nonlinear
parameters}(c) show the temperature dependence of the AO
parameters in the regimes A and B discussed above:
temperature-dependent $Q_{\rm A}$, $I_{\rm th,A}$, $I_{\rm MI}$,
and temperature-independent $Q_{\rm B}$,$I_{\rm th,B}$. The larger
value of $Q_{\rm B}$ relative to $Q_{\rm A}$ observed in Fig.
\ref{fig:nonlinear parameters}(b), indicates that the mode
couplings in regime B cause additional nonlinear damping compared
to regime A. In regime A where a single mode C1 is dominant,
$Q_{\rm A}$ depends on temperature such that it is almost constant
at $T \leq 125$ K but increases for $T >$ 125 K. This reflects the
temperature dependence of the minimum linewidth and maximum power
of C1 in Fig. \ref{fig:nonlinear parameters}(g) and
\ref{fig:nonlinear parameters}(h) and the existence of another
temperature-independent relaxation mechanism below 125 K for an
individual AO mode. Also, as $T$ increases, both $I_{\rm th,A}$
and $I_{\rm MI}$ decrease linearly, demonstrating that thermal
fluctuation noise facilitates the generation and stabilization of
the AO mode and of mode-mode coupling with the smaller $I_{\rm
DC}$ such as
\begin{align}
I_{\rm th} \left( T \right) &= I_{\rm th,0} + \kappa_{\rm th} T \label{eqn:Ith}\\
I_{\rm MI} \left( T \right) &= I_{\rm MI,0} + \kappa_{\rm MI} T
\label{eqn:IMI}
\end{align}
where $I_{\rm th,0}$ and $I_{\rm MI,0}$ are the intrinsic
threshold and transition currents with thermal fluctuation
excitation effects removed, and $\kappa_{\rm th}$ and $\kappa_{\rm
MI}$ are coefficients with respect to temperature change. The
temperature dependence of the intrinsic threshold, $I_{\rm th,0}
\left( T \right)$, can be obtained from a separately measured spin
torque FMR, where $I_{\rm th,0}$ is estimated to be almost
constant in the temperature range of 5 - 300 K (See Appendix
\ref{section:intrinsicIth}).

There are significant changes in the current evolution of the AO
modes when $\theta$ changes from $65^{\circ}$ to $70^{\circ}$ as
shown in Fig. \ref{fig:fmr parameters}(a) and \ref{fig:fmr
parameters}(f). Edge-bulk combined modes (C1, C2, C3) at $\theta =
65^{\circ}$ (see Fig. \ref{fig:spectrum}(e) and Fig. \ref{fig:fmr
parameters}(a)-(e)) are spectrally and spatially separated into
edge (E1, E2) and bulk (B1, B2, B3) modes at $\theta = 70^{\circ}$
(see Fig. \ref{fig:spectrum}(d) and Fig. \ref{fig:fmr
parameters}(f)-(j)). As a result, for the AO modes existing at
$\theta = 70^{\circ}$ (see Fig. \ref{fig:fmr parameters}(g)-(i)),
the evolution of $f_0$, $\Delta f$, and $P$ of the AO modes with
increasing $I_{\rm DC}$ are more complex than for $65^{\circ}$:
the E1 and E2 modes split at 7 mA, and there can be various
mode-mode couplings with different strengths of dipolar and
exchange interactions depending on the spatial distance between
the two interacting modes (e.g., edge-edge, edge-bulk and
bulk-bulk). This information can be valuable in developing
strategies that employ spectral and spatial mode separation to
reduce mode couplings and thus enhance the performance of
auto-oscillators. We note that these mode couplings should be
distinguishable from the nonlinear bullet mode dynamics. The power
sum of E1 and E2 monotonically increases up to $I_{\rm MI} = 7.2$
mA until it saturates creating a kink in $P_{\rm total}$ as shown
in Fig. \ref{fig:fmr parameters}(i). The temperature dependence of
the AO parameters for $\theta = 70^{\circ}$ shown in Fig.
\ref{fig:nonlinear parameters}(e) and \ref{fig:nonlinear
parameters}(f) is generally similar to that for $\theta =
65^{\circ}$, except for a further increased $Q_{B}$ value, perhaps
due to the increased number of intermode interaction routes
allowed between a larger number of AO modes.

In conclusion, we observe the nonlinear damping transition of the
auto-oscillation modes through the total power as a function of
drive current, which coincides with the onset of power
distribution amongst multiple modes and linewidth broadening of
all individual modes. We find the nonlinear damping due to
mode-mode coupling to be independent of temperature, which
suggests that the mode coupling occurs through intermode
interactions such as dipole and exchange interactions rather than
thermally excited magnon mediated nonlinear scattering. This study
of nonlinear damping due to mode couplings presents a promising
pathway to overcome the current limitations of efficiently
controlling mode interactions in spin Hall nano-oscillators to
prevent mode crosstalk or inter-mode energy transfer and deepens
understanding of complex nonlinear spin dynamics in multimode spin
wave systems. As one solution, we can generate well-defined AO
modes locally excited by the dipole field from a nano- or
micron-scale permanent magnet \cite{Zhang2021}, where the number
of modes and their frequency distribution can be tuned by changing
the local dipole field by moving a permanent magnet relative to
the sample surface. Furthermore, the interaction between two
spatially separated AOs can be controlled by adjusting the
relative lateral distance in a scanned system
\cite{Lee2010,Lee2011,Zhang2019,Zhang2021}.

\begin{acknowledgments}
We thank Denis V. Pelekhov for helpful discussions. This work was
primarily supported by the Center for Emergent Materials, an NSF
MRSEC, grant DMR-2011876.
\end{acknowledgments}

\appendix

\section{Angle Dependence of auto-oscillation spectrum over
extended angle range}\label{section:extendspectrum}

\begin{figure}
\centering
\includegraphics[width=1.0\columnwidth]{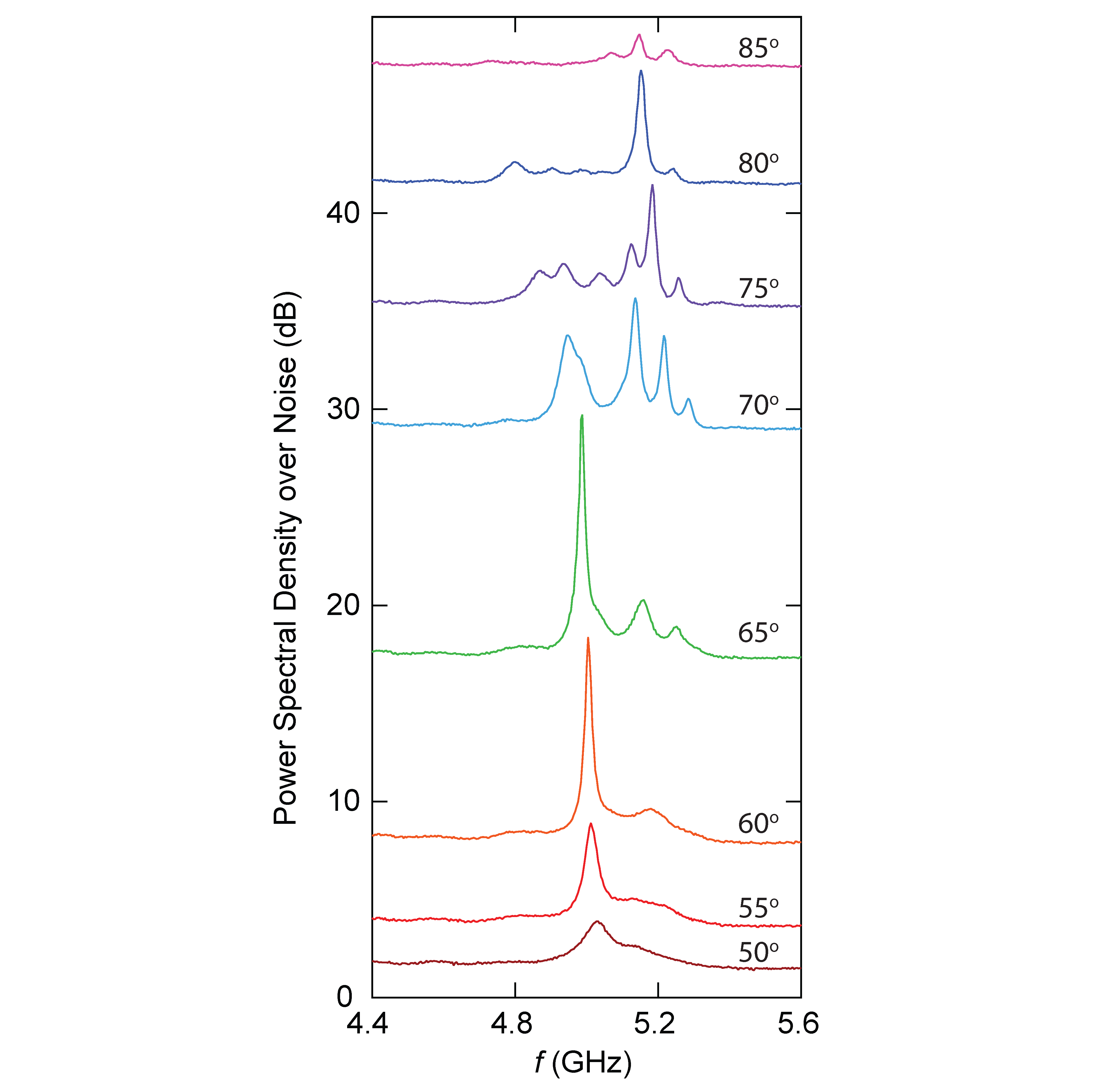}
\caption{Angle dependence of the auto-oscillation spectrum in Fig.
\ref{fig:spectrum}(a) of the main text extended to a wide angle
range. It is measured at $I_{\rm DC} = 7$ mA, $T = 77$ K and
$H_{0} = 570$ Oe.} \label{fig:AO_spectrum_ang_dep}
\end{figure}

Fig. \ref{fig:AO_spectrum_ang_dep} shows the dependence of the
auto-oscillation spectrum on angle, similar to Fig.
\ref{fig:spectrum}(a), but over an extended angular range: $\theta
= 50^{\circ} - 85^{\circ}$. At $\theta \leq 65^{\circ}$, the
lowest frequency mode has the largest amplitude and shifts
slightly towards lower frequencies as $\theta$ increases. The mode
starts to split and its amplitude decreases above $70^{\circ}$
while the frequency shifts monotonically lower with increasing
$\theta$. On the other hand, the high frequency modes hardly shift
with increasing $\theta$, and at $\theta \geq 70^{\circ}$, one of
them has the largest amplitude among all excited AO modes in the
spectrum instead of the lowest frequency mode. Our micromagnetic
simulations in the next section show that edge modes at low
frequencies shift monotonically to lower frequencies, while bulk
modes at high frequencies hardly shift; this characteristic
behavior allows edge and bulk modes to be differentiated for
$\theta \geq 70^{\circ}$.

\section{Micromagnetic Simulations}\label{section:simulation}

We perform micromagnetic simulations using MuMax3 \cite{Mumax3} to
understand complex AO spectra and identify their relevant spatial
mode profiles. In simulations, the computational dimension 2.6
$\mu$m $\times$ 9 $\mu$m $\times$ 5 nm is subdivided into 5 nm
$\times$ 17.5 nm $\times$ 5 nm cells. As magnetic parameters, we
use the gyromagnetic ratio of $\gamma / 2 \pi =$ 2.8 MHz/G and
effective magnetization $4 \pi M_{\rm eff} =$ 6502 G obtained from
the ST-FMR data (see Appendix \ref{section:STFMR}) measured at
$I_{\rm DC} =$ 0 via Kittel equation.
\begin{equation}
f_0 = \frac{\gamma}{2 \pi} \left[ H_0 \left( H_0 + 4 \pi M_{\rm
eff} \right) \right]^{1 / 2}
\end{equation}
The standard values of exchange stiffness $A_{\rm ex} = 1.3 \times
10^{-11}$ J/m and the Gilbert damping constant $\alpha =$ 0.01 for
permalloy are used.

\subsection{Linear Spin Wave Eigenmode}

First, we perform micromagnetic simulations in the linear regime
of magnetodynamics with no bias current. Initially, the magnetic
system is excited by a sinc rf field with an amplitude of 10 mT
and a cutoff frequency of 40 GHz. Then Gilbert damping is turned
off by making $\alpha = 0$, allowing the magnetic dynamic system
to proceed freely for 187 ns. We obtain the spectral and spatial
profiles of the modes by performing Fourier transform with dynamic
motion after 62 ns to avoid initial transients \cite{Dvornik2018}.

\begin{figure*}
\centering
\includegraphics[width=2\columnwidth]{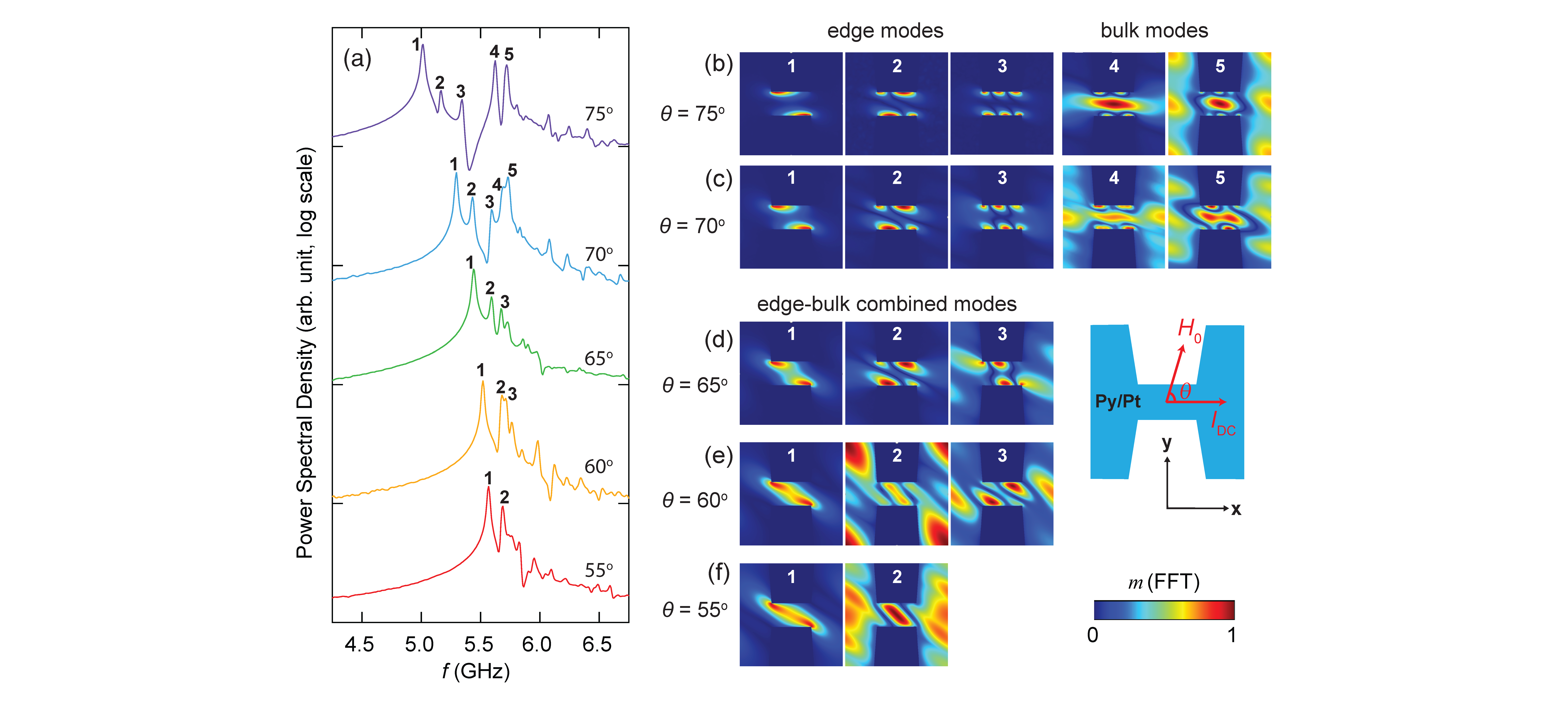}
\caption{(a) Spectra for various $\theta$ obtained from the
micromagnetic simulation using Mumax3 with zero bias current,
which reflect the linear spin wave eigenmodes determined by the
applied field $\mathbf{H_{0}}$ and the constricted sample
geometry. This is in contrast to the case of nonlinear
self-oscillatory modes with relatively larger amplitudes driven by
the applied anti-damping spin torque. In (a), each spectrum is
vertically offset. (b)-(f) Spatial eigenmode profiles
corresponding to the spectral peaks indicated by the numbers in
(a) at various $\theta$. The dynamic magnetization amplitude $m$
is normalized on the color scale of each image. At
$\theta=70^{\circ}$ and $75^{\circ}$, the spatial separation of
edge (1-3) modes and bulk (4-5) modes occurs as shown in (b) and
(c) as well as their spectral separation as shown in (a).}
\label{fig:linear_spin_wave_mode}
\end{figure*}

Fig. \ref{fig:linear_spin_wave_mode}(b)-(f) show the spatial
profile of the linear spin wave eigenmodes corresponding to each
spectral peak indicated by the number in Fig.
\ref{fig:linear_spin_wave_mode}(a) for each angle $\theta$. The
edge and bulk modes are combined for $\theta \leq 65^{\circ}$,
whereas the edge and bulk modes are spatially and spectrally
separated for $\theta \geq 70^{\circ}$ due to the significant
difference in the demagnetizing field between the edge and center
regions. Edge modes at low frequencies shift monotonically to
lower frequencies, and bulk modes at high frequencies shift
little.

\subsection{Self-Oscillatory Mode}

The auto-oscillation mode of the system is obtained by solving the
Landau-Lifshitz-Gilbert equation with an anti-damping spin torque
applied to the active region of the AO device. In the simulations,
we try two different current density distributions shown in Fig.
\ref{fig:J_HOe_comsol} and Fig. \ref{fig:J_HOe_gaus} considered
for a 1 mA bias current.

As an initial state in the simulation, the magnetic system is
allowed to relax to a state close to the ground state. The
anti-damping torque proportional to the current density in Fig.
\ref{fig:J_HOe_comsol} and \ref{fig:J_HOe_gaus} scaled by the
current value is activated at 0 ns. If this anti-damping torque is
smaller than the Gilbert damping, the magnetization oscillations
decay, whereas if the anti-damping torque is larger than the
Gilbert damping, the magnetization oscillations grow until their
amplitude saturates. We define a threshold current as that at
which anti-damping is balanced with Gilbert damping, where the
magnetization oscillates with a constant amplitude over time. The
scale factor for the current in the simulations is chosen so that
the threshold currents in the simulations are as close as possible
to the threshold currents in the actual experimental data. The
spectral and spatial profiles of the excitation modes are obtained
by performing Fourier transforms of the time dependence of the
magnetization dynamics.

\subsubsection{Micromagnetic simulation using current density
distribution calculated in COMSOL}

\begin{figure}
\centering
\includegraphics[width=1\columnwidth]{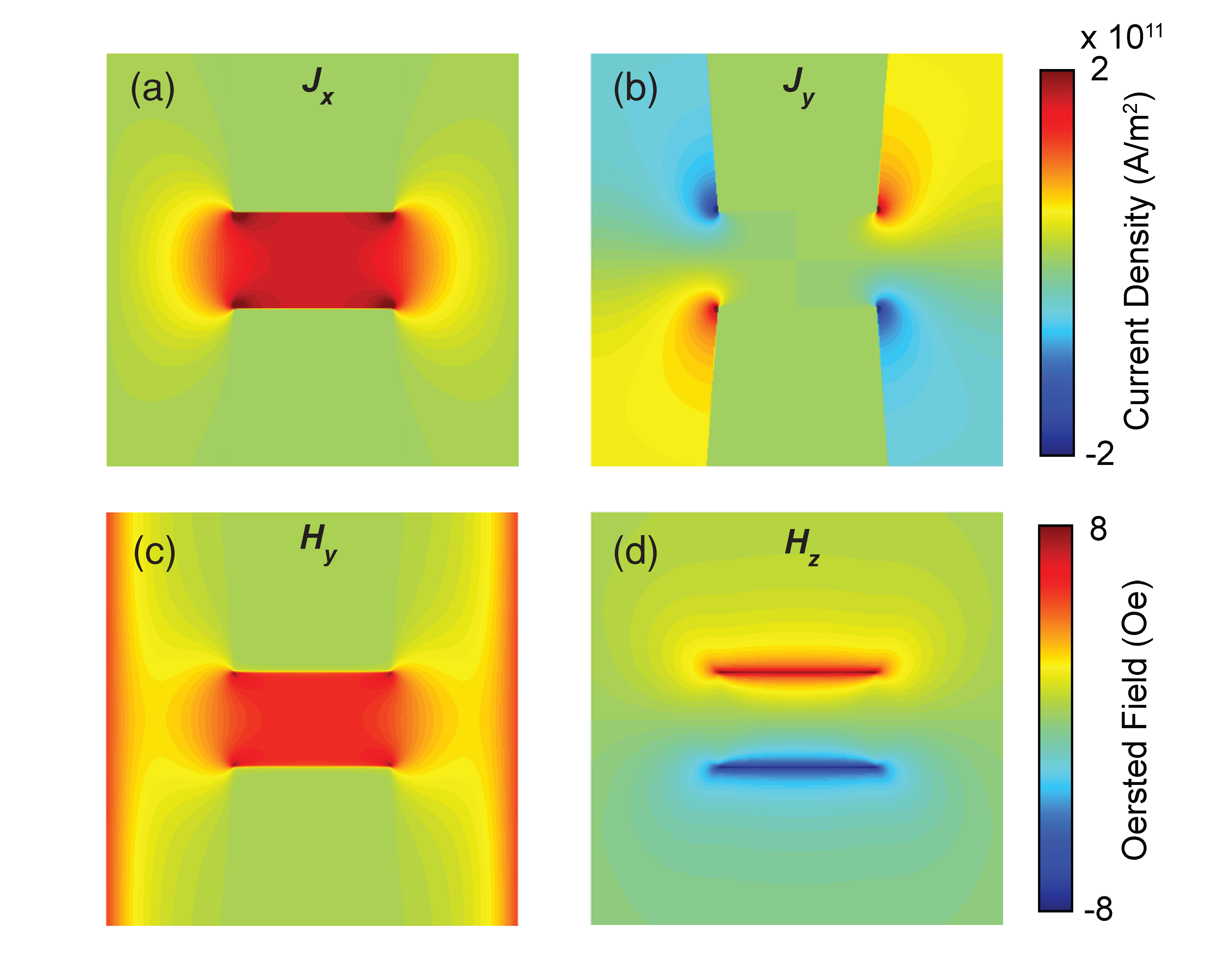}
\caption{COMSOL \cite{comsol} calculation of current densities in
the AO structure for a 1 mA current: (a) x-component of current
density distribution $J_{\rm x}$, (b) y-component of current
density distribution $J_{\rm y}$, (c) y-component of Oersted field
$H_{\rm y}$, and (d) z-component of Oersted field $H_{\rm z}$.
$J_{\rm z} = 0$ and $H_{\rm x} = 0$. Anti-damping spin torque in
the micromagnetic simulations is calculated based on these maps.
The corresponding results are shown in Fig.
\ref{fig:currdep_simu_comsol} and Fig.
\ref{fig:angledep_simu_comsol}.} \label{fig:J_HOe_comsol}
\end{figure}

\begin{figure}
\centering
\includegraphics[width=1\columnwidth]{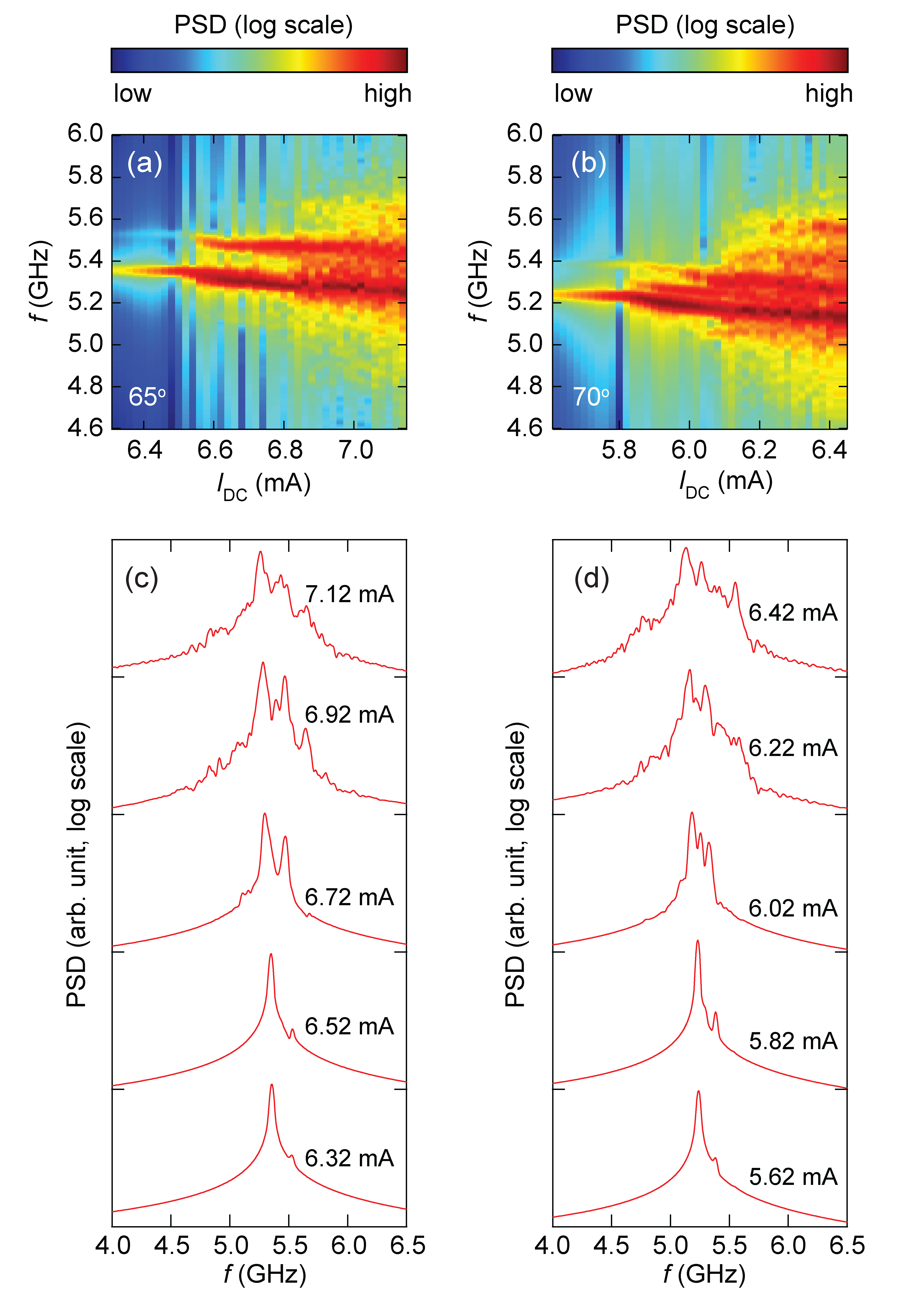}
\caption{The evolution of power spectral density (PSD) with
increasing bias current $I_{\rm DC}$ for (a) $\theta=65^{\circ}$
and (b) $\theta=70^{\circ}$ obtained from micromagnetic
simulations performed with the current density distribution and
Oersted field in Fig. \ref{fig:J_HOe_comsol}. (c) Representative
spectrum linecuts in (a) for various $I_{\rm DC}$ at
$\theta=65^{\circ}$. (d) Representative spectral linecuts in (b)
for various $I_{\rm DC}$ at $\theta=70^{\circ}$.}
\label{fig:currdep_simu_comsol}
\end{figure}

\begin{figure*}
\centering
\includegraphics[width=2\columnwidth]{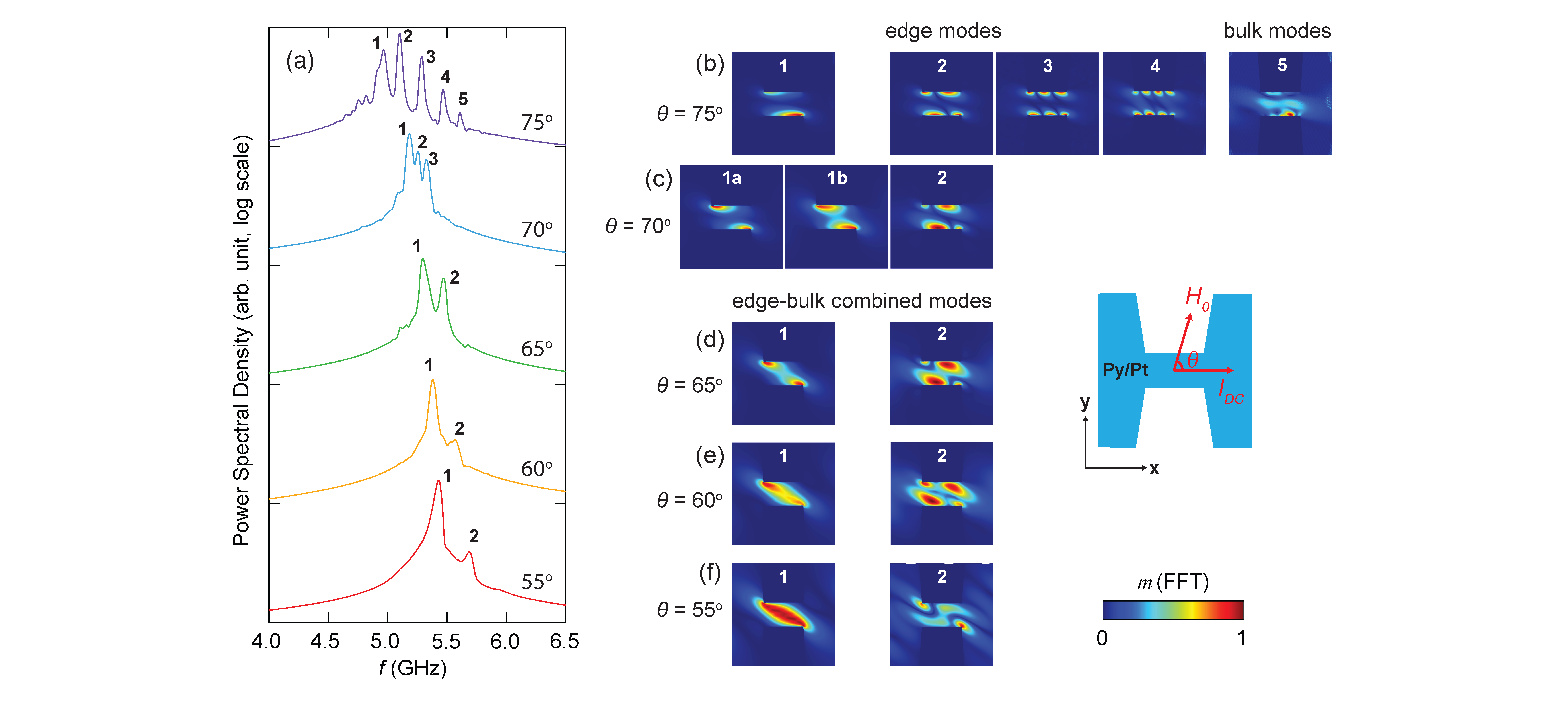}
\caption{(a) Spectra for various $\theta$ obtained from
micromagnetic simulations using Mumax3 performed with the current
density distribution and Oersted field in Fig.
\ref{fig:J_HOe_comsol}. The corresponding $I_{\rm DC}$ are 8.28 mA
for $\theta=55^{\circ}$, 7.42 mA for $\theta=60^{\circ}$, 6.72 mA
for $\theta=65^{\circ}$, 6.02 mA for $\theta=70^{\circ}$, and 5.58
mA for $\theta=75^{\circ}$, 0.2 mA larger than the respective
threshold current at each $\theta$. Each spectrum is vertically
offset for clarity. (b)-(f) Spatial eigenmode profiles
corresponding to the spectral peaks indicated by numbers in (a) at
various $\theta$. The dynamic magnetization amplitude $m$ is
normalized on the color scale of each image. At
$\theta=70^{\circ}$ and $75^{\circ}$, the spatial separation of
edge (1-4) and bulk (5) modes occurs as shown in (b) and (c) as
well as their spectral separation as shown in (a). Compared to the
linear spin wave modes in Fig. \ref{fig:linear_spin_wave_mode},
the bulk modes are mostly suppressed in this simulation.}
\label{fig:angledep_simu_comsol}
\end{figure*}

Fig. \ref{fig:J_HOe_comsol} shows the current density distribution
of the AO system for a 1 mA current calculated using COMSOL
\cite{comsol} and the Oersted field produced by it. These
determine the anti-damping torque in micromagnetic simulations,
and the corresponding results are shown in Fig.
\ref{fig:currdep_simu_comsol} and Fig.
\ref{fig:angledep_simu_comsol}.

Figs. \ref{fig:currdep_simu_comsol}(a) and
\ref{fig:currdep_simu_comsol}(b) show the evolution of power
spectral density (PSD) with increasing bias current $I_{\rm DC}$
for $\theta=65^{\circ}$ and $\theta=70^{\circ}$, respectively, and
their representative linecuts are shown in Fig.
\ref{fig:currdep_simu_comsol}(c) and
\ref{fig:currdep_simu_comsol}(d), respectively. These simulation
data show the key features of the experimental data in Fig.
\ref{fig:fmr parameters} of the main text, such as monotonic
redshift of the resonant frequency, mode amplitude growing and
mode amplitude saturation with increasing $I_{\rm DC}$. In the
simulation, the strong mode broadening starts at a relatively
lower $I_{\rm DC}$ compared to the experimental data, so Fig.
\ref{fig:currdep_simu_comsol} shows the simulation result
applicable only to the low current region of the experimental data
Fig. \ref{fig:fmr parameters}. This indicates that the nonlinear
magnonic effect occurring at high currents, which can cause
stronger self-excitation of the AO mode with high coherence, seems
to be lacking in the simulations. Since thermal effects play no
role in the simulations, the mode broadening seen at high $I_{\rm
DC}$ arises purely from dipole or exchange spin-spin interactions
of dynamic magnetization occurring at relatively large cone
angles. This directly demonstrates nonlinear damping due to mode
couplings that occur via dipole or exchange interactions,
supporting the main conclusion of the paper.

Fig. \ref{fig:angledep_simu_comsol}(a) shows the spectrum for
various $\theta$ and Fig. \ref{fig:angledep_simu_comsol}(b)-(f)
show the spatial eigenmode profiles corresponding to the spectral
peaks indicated by the numbers in Fig.
\ref{fig:angledep_simu_comsol}(a). Most of the AO modes excited by
the anti-damping torque are edge modes that shift only to lower
frequencies with increasing $\theta$, while the barely shifted
bulk modes shown in the linear spin wave modes in Fig.
\ref{fig:linear_spin_wave_mode} are mostly suppressed. Compared
with the spatial mode profiles of the linear spin wave modes in
Fig. \ref{fig:linear_spin_wave_mode}, the edge modes in this
simulation seems to be selectively excited and evolved from the
linear spin wave modes by the inhomogeneous current density
distribution.

\subsubsection{Micromagnetic simulation using current density distribution of
2D Gaussian model}

\begin{figure}
\centering
\includegraphics[width=1\columnwidth]{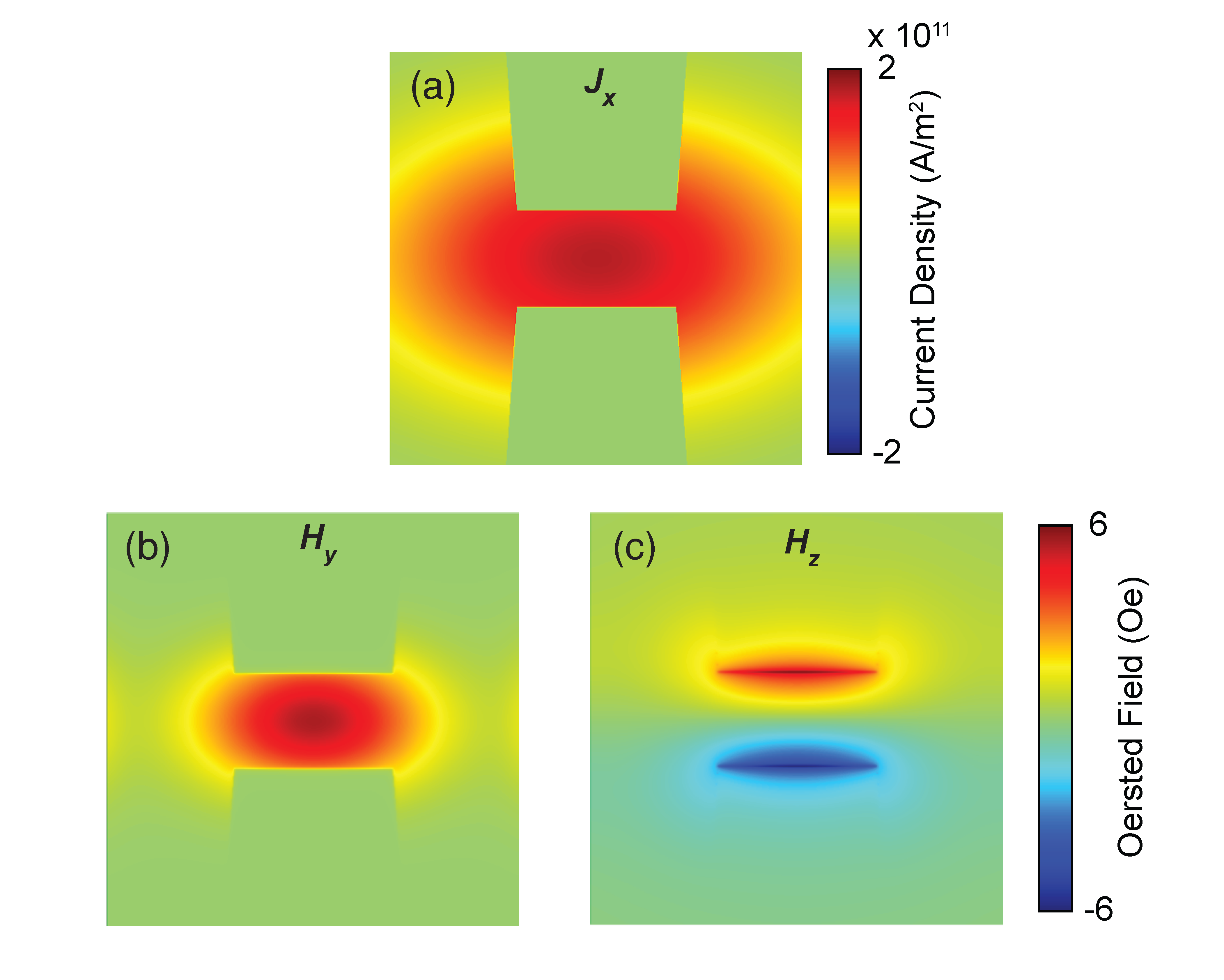}
\caption{(a) x-component of current density distribution $J_{\rm
x}$, (b) y-component of Oersted field $H_{\rm y}$, and (c)
z-component of the Oersted field $H_{\rm z}$ in the AO system for
1 mA current used to apply anti-damping spin torque in the
micromagnetic simulations. $J_{\rm y} = J_{\rm z} = 0$ and $H_{\rm
x} = 0$. The corresponding results are shown in Fig.
\ref{fig:spectrum} and Fig. \ref{fig:currdep_simu_comsol}.}
\label{fig:J_HOe_gaus}
\end{figure}

\begin{figure}
\centering
\includegraphics[width=1\columnwidth]{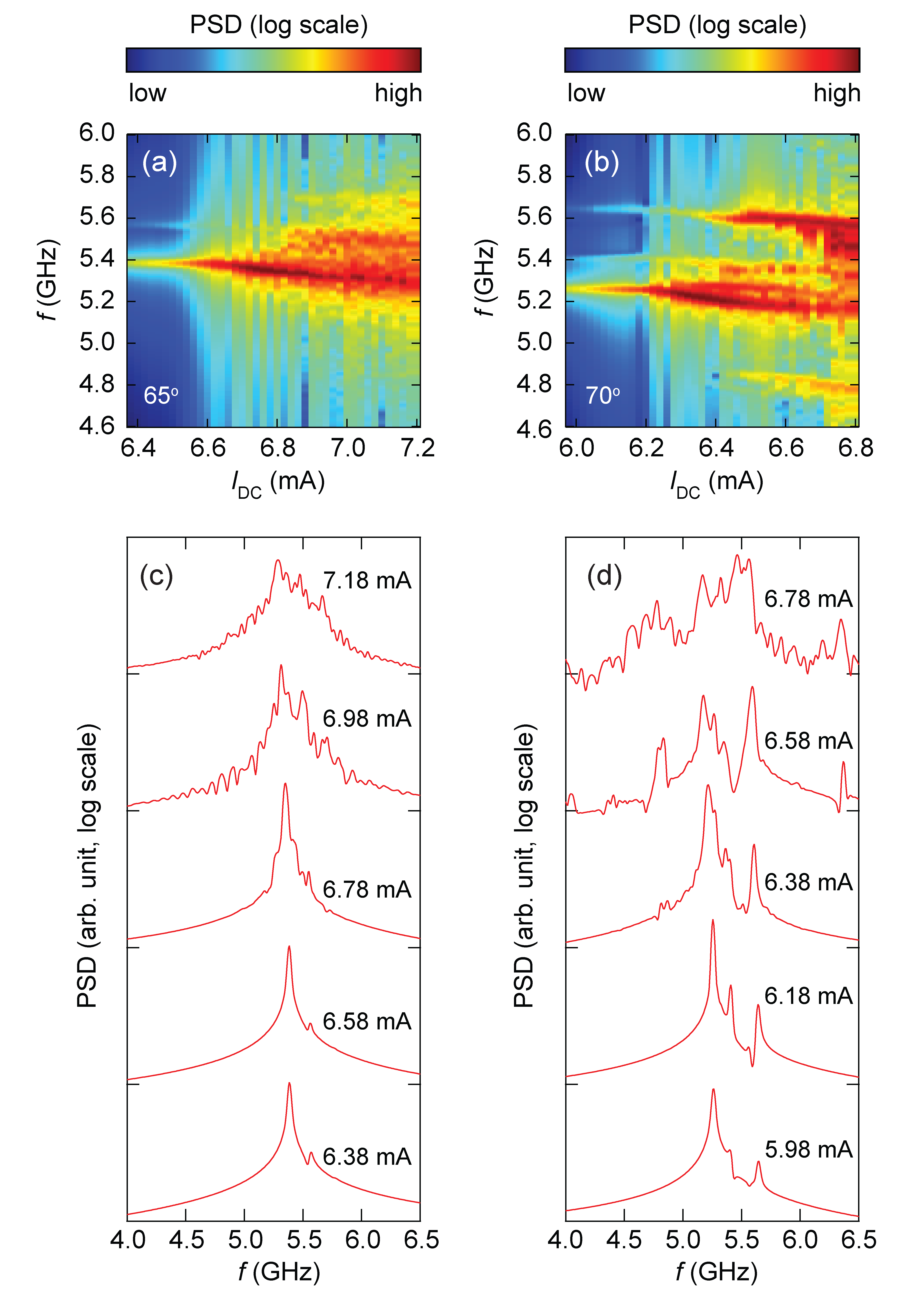}
\caption{The evolution of power spectral density (PSD) with
increasing bias current $I_{\rm DC}$ for (a) $\theta=65^{\circ}$
and (b) $\theta=70^{\circ}$ obtained from micromagnetic
simulations performed with the current density distribution and
Oersted field in Fig. \ref{fig:J_HOe_gaus}. (c) Representative
spectrum linecuts in (a) for various $I_{\rm DC}$ at
$\theta=65^{\circ}$. (d) Representative spectrum linecuts in (b)
for various $I_{\rm DC}$ at $\theta=70^{\circ}$.}
\label{fig:currdep_simu_gaus}
\end{figure}

In order to test the effect of the inhomogeneous current density
distribution on the excitation of the AO modes in the simulation,
we try a micromagnetic simulation using the different current
density distribution in Fig. \ref{fig:J_HOe_gaus}(a), where the
current density $J_{\rm x}$ has a 2D Gaussian distribution with a
maximum at the center and a broader distribution as function of
position in the device compared to Fig. \ref{fig:J_HOe_comsol}(a),
and $J_{\rm y} = J_{\rm z} = 0$. Indeed, the spectral shapes in
this simulation are considerably different as shown in Fig.
\ref{fig:spectrum}(b) of the main text and Fig.
\ref{fig:currdep_simu_gaus}(c) and \ref{fig:currdep_simu_gaus}(d).
In contrast to Fig. \ref{fig:angledep_simu_comsol}, bulk modes
excited around the central region are unsuppressed for
$\theta=70^{\circ}$ and $75^{\circ}$ and can have larger mode
amplitudes than edge modes, as shown in Fig. \ref{fig:spectrum}.
These bulk modes exhibit improved agreement with experimentally
observed AO modes, in particular small frequency shift with
increasing $\theta$ as shown in Fig. \ref{fig:spectrum}(a) and
Fig. \ref{fig:AO_spectrum_ang_dep}. Also, the overall spectral
shapes associated with the number of the excited AO modes, their
relative frequencies and amplitudes, their frequency shift
behaviors for varying $\theta$ in Fig. \ref{fig:spectrum}(b)
better match the experimental data in Fig. \ref{fig:spectrum}(a)
compared to the other AO simulations in Fig.
\ref{fig:linear_spin_wave_mode}(a) and Fig.
\ref{fig:angledep_simu_comsol}(a). Therefore, we conclude that the
actual current density distribution is closer to the 2D Gaussian
function in Fig. \ref{fig:J_HOe_gaus} than that calculated using
COMSOL in Fig. \ref{fig:J_HOe_comsol}. We note that unidentified
structural defects around the edges of the AO active region of the
sample can significantly alter the spectral shapes, especially for
$\theta=70^{\circ}$ and $75^{\circ}$, where the mode constriction
effect is stronger.

This conclusion is further supported by the evolution of the power
spectral density (PSD) with increasing bias current $I_{\rm DC}$
for $\theta=65^{\circ}$ and $70^{\circ}$ shown in Fig.
\ref{fig:currdep_simu_gaus}(a) and \ref{fig:currdep_simu_gaus}(b),
respectively. In Fig. \ref{fig:currdep_simu_gaus}(a) and
\ref{fig:currdep_simu_gaus}(c) for $\theta=65^{\circ}$, the edge
mode at the lowest frequency has the largest amplitude at any
$I_{\rm DC}$ whereas in Fig. \ref{fig:currdep_simu_gaus}(b) and
\ref{fig:currdep_simu_gaus}(d) for $\theta=70^{\circ}$, the bulk
mode at $\sim$ 5.6 GHz grows with increasing $I_{\rm DC}$ and
eventually has the larger amplitude than any edge modes located at
the lower frequencies at high $I_{\rm DC}$, which agrees well with
our experimental data in Fig. \ref{fig:fmr parameters}.

\section{Quantifying Nonlinear Damping with the Coefficient
$Q$}\label{section:Q}

The nonlinear single-mode auto-oscillator can be described with a
universal oscillator model, another form of the Landau-Lifshitz
equation, derived by Slavin and Tiberkevich \cite{Slavin2009},
\begin{equation}
\frac{dc}{dt} + i \omega (p) c + \Gamma_{+} (p) c - \Gamma_{-} (p)
c = f_{n} (t) \label{eqn:oscillator model}
\end{equation}
where $c$ is the complex dimensionless dynamic magnetization
amplitude, $p\, (= |c|^2)$ is the oscillation power, $\omega (p)$
is the power-dependent nonlinear frequency,
\begin{align}
\Gamma_{+} (p) &\approx \Gamma_{G} ( 1 + Q p )\label{eqn:Gammapos}\\
\Gamma_{-} (p) &\approx \sigma I ( 1 - p )\label{eqn:Gammaneg}
\end{align}
are the positive and negative damping constant, respectively,
$f_{n} (t)$ is the thermal fluctuation noise, $\Gamma_{G} \left( =
\alpha \omega \right)$ is the Gilbert damping, $Q$ is the
nonlinear damping coefficient, $\sigma$ is the spin-current
efficiency, and $I$ is the drive current. On the left side of Eq.
(\ref{eqn:oscillator model}), the second term describes
precession, the third term describes damping, and the fourth term
describes anti-damping. Note that $\omega (p)$, $\Gamma_{+} (p)$,
and $\Gamma_{-} (p)$ are auto-oscillation power $p$ dependent.

\begin{figure}
\centering
\includegraphics[width=1\columnwidth]{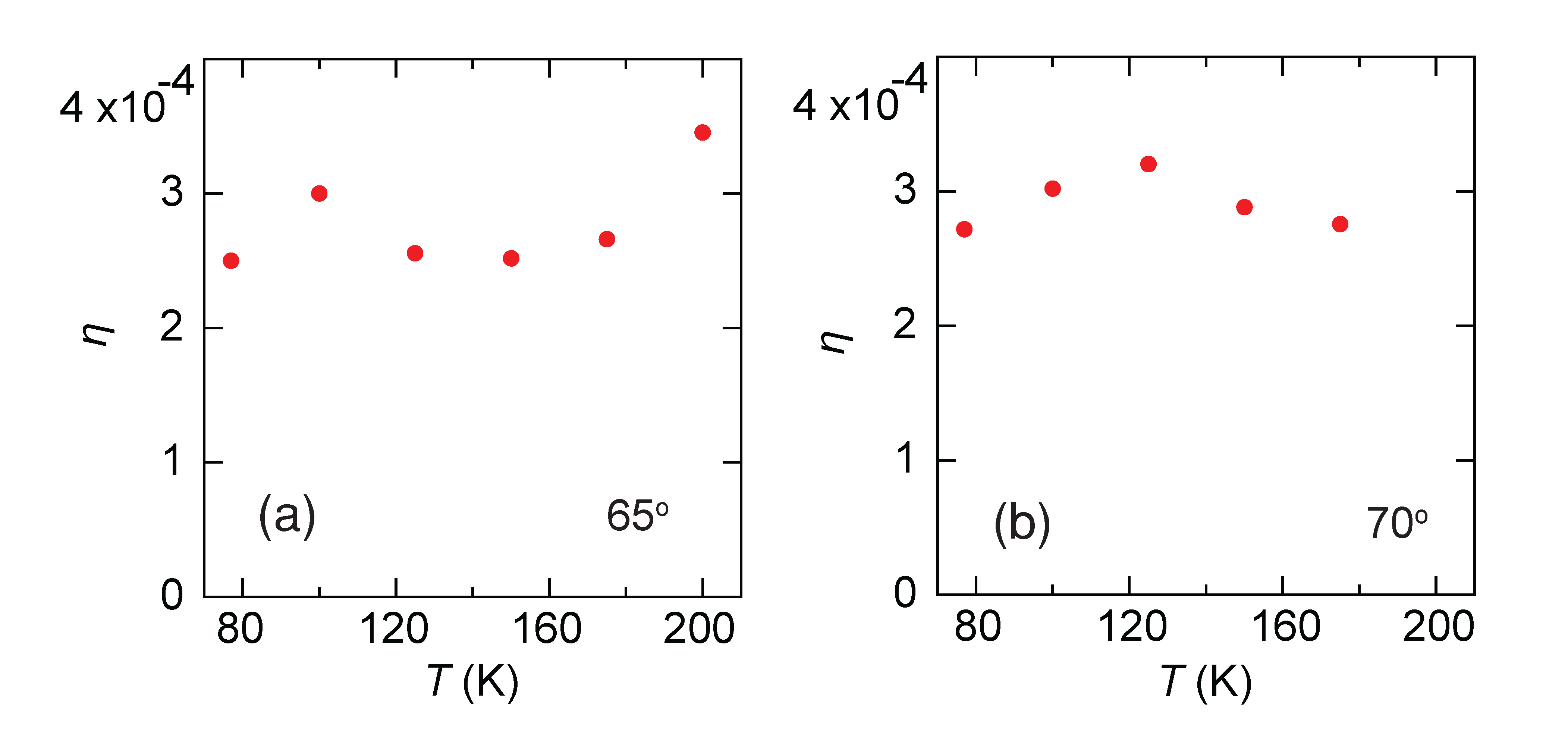}
\caption{Noise level $\eta$ vs. temperature $T$ for (a) $\theta =
65^{\circ}$ and (b) $\theta = 70^{\circ}$ obtained from the
fitting curves of solid colored lines in Fig. \ref{fig:nonlinear
parameters}(a) and \ref{fig:nonlinear parameters}(d).}
\label{fig:noise vs T}.
\end{figure}

In order to describe the nonlinear damping of the multimode AO
system in a simple and quantitative way, we calculate $Q$ from the
relationship with $P_{\rm total} = N_{0} p$, where for $I_{\rm DC}
< I_{\rm MI}$,
\begin{equation}
p = \frac{Q \eta}{Q + \eta} \left[ 1 + \frac{\exp(-(\zeta + Q) /
Q^2 \eta)}{E_{\beta}((\zeta + Q)/Q^2 \eta)} \right] + \frac{\zeta
- 1}{\zeta + Q} \label{eqn:PtotallIMI}
\end{equation}
for $I_{\rm DC} \geq I_{\rm MI}$,
\begin{equation}
p = \frac{\zeta - 1}{\zeta + Q} \label{eqn:PtotalgeqIMI}
\end{equation}
In these expressions, $N_{0}$ is the power conversion factor for
the normalized $p$, $\zeta = I/I_{\rm th}$, $I_{\rm th}$ is the
threshold current, $\eta$ is the effective noise power, $E_{\rm
\beta} (x) = \int_{1}^{\rm \infty} e^{-xt} / t^{\beta} dt$, and
$\beta = -(1 + Q) \zeta / Q^2 \eta$ \cite{Slavin2009}. Note that
the Slavin-Tiberkevich model was originally intended to describe
single-mode auto-oscillators, so $Q$ in the multimode excitation
regime of $I_{\rm DC} \geq I_{\rm MI}$ is an  \textquotedblleft
effective\textquotedblright nonlinear damping parameter of the
entire AO system. We also obtain $\eta$ in Eq.
(\ref{eqn:PtotallIMI}) along with other AO parameters presented in
Fig. \ref{fig:nonlinear parameters} by fitting, but Fig.
\ref{fig:noise vs T} reveals no credible temperature dependence to
$\eta$.

\section{Spin Torque Ferromagnetic Resonance}\label{section:STFMR}

\begin{figure}
\centering
\includegraphics[width=1\columnwidth]{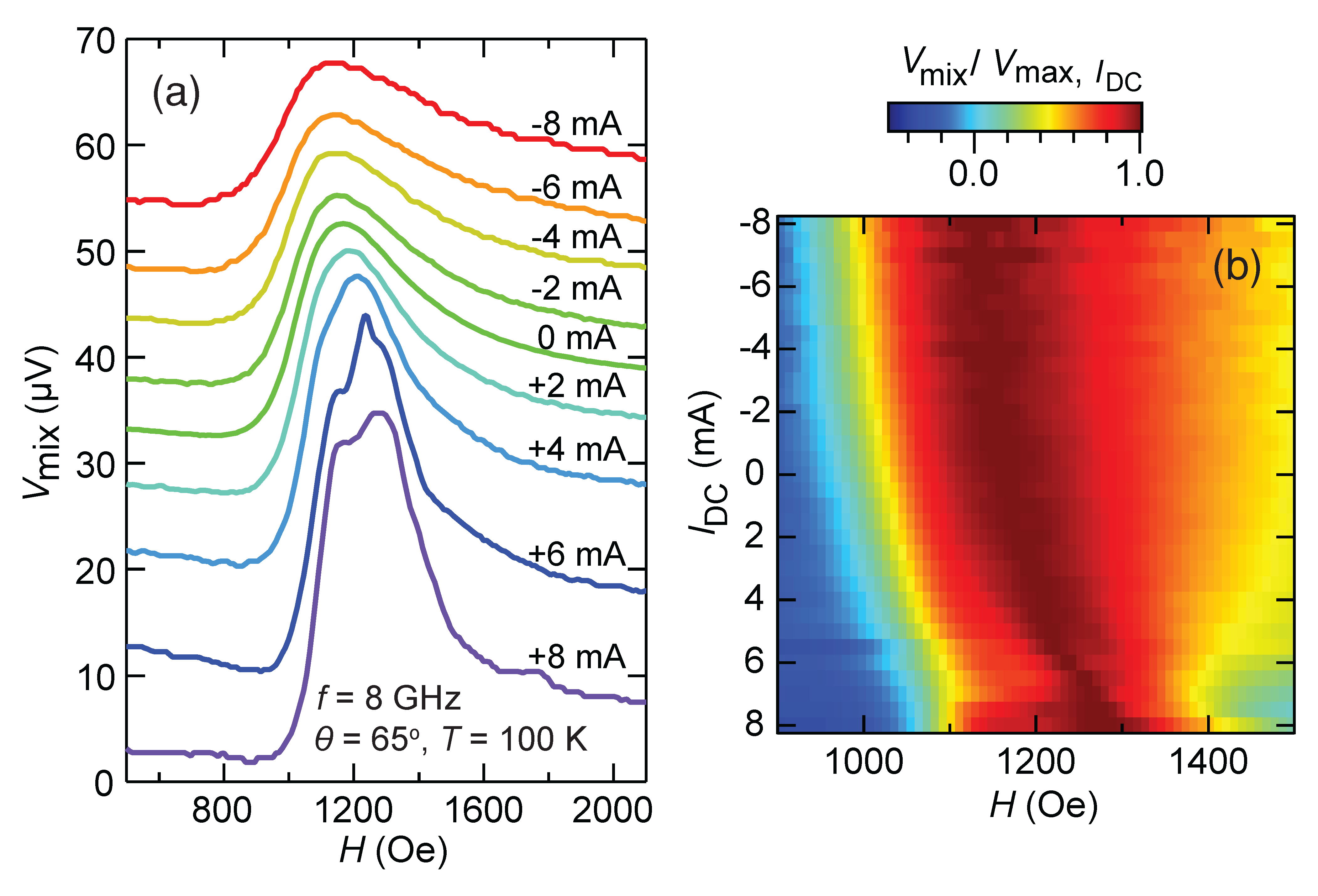}
\caption{(a) Spin torque FMR (ST-FMR) spectrum at various $I_{\rm
DC}$ ranging from -8 mA to 8 mA. (b) The DC current evolution of
ST-FMR where the signal $V_{\rm mix}$ is normalized to the maximum
at each $I_{\rm DC}$ to see the shift of ST-FMR peak. Obviously,
the monotonic ST-FMR peak shift to the lower field with no change
in the shift to the higher field for the intensity of $I_{\rm DC}
< 0$ increasing up to -8 mA indicates no reduction in saturation
magnetization that can be caused by Joule heating at large
intensity of $I_{\rm DC}$.} \label{fig:IDC Dep of STFMR}
\end{figure}

Fig. \ref{fig:IDC Dep of STFMR} (a) shows a typical spin torque
ferromagnetic resonance (ST-FMR) spectrum measured at various DC
currents $I_{\rm DC}$ ranging from -8 mA to 8 mA. As the absolute
value of $I_{\rm DC} < 0$ increases, the ST-FMR peak broadens due
to the enhanced damping, while as $I_{\rm DC} > 0$ increases, the
mode becomes narrower due to anti-damping and eventually splits
into a couple of modes. In order to see the overall shift of the
ST-FMR peak, we normalize the signal $V_{\rm mix}$ to the maximum
at each $I_{\rm DC}$, which is shown in Fig. \ref{fig:IDC Dep of
STFMR} (b). With increasing $I_{\rm DC} > 0$, the ST-FMR peak
shifts to the higher field because as the amplitudes of
self-localized AO modes grow, the static magnetization decreases.
On the other hand, as the absolute value of $I_{\rm DC} < 0$
increases up to -8 mA, the ST-FMR peak shifts monotonically to the
lower field and does not show a shift to the higher field that
might occur due to the decrease in saturation magnetization by
Joule heating. This means that even at $I_{\rm DC} = 8$ mA
regardless of its sign, our device's magnetic system has not
reached the current regime where the saturation magnetization
decreases by Joule heating \cite{Demidov2011}. This provides
crucial evidence that the contribution of Joule heating to the
nonlinear damping of the AO system described in the main text is
not significant at high $I_{\rm DC}$ up to 8 mA.

\begin{figure}
\centering
\includegraphics[width=1\columnwidth]{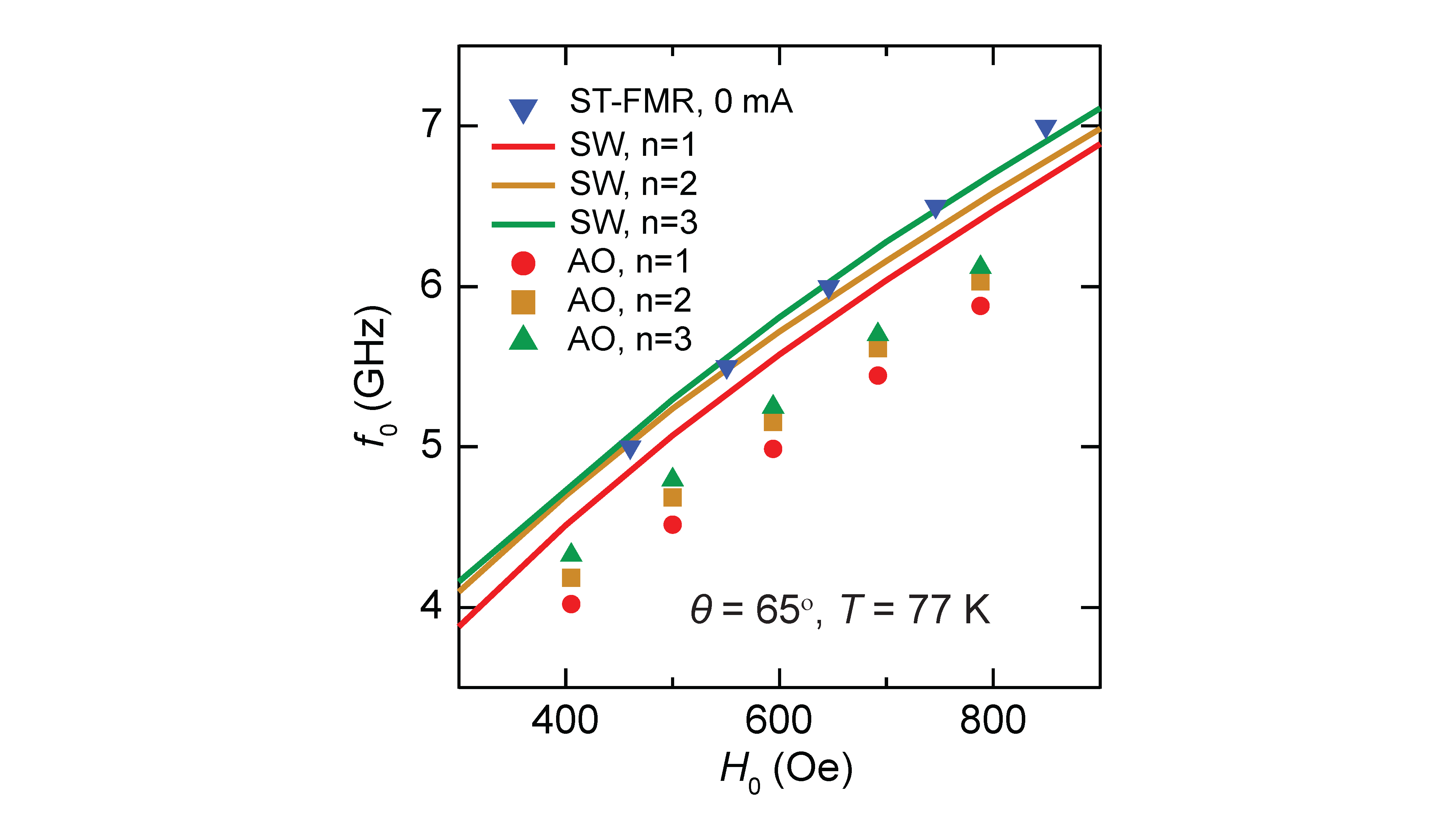}
\caption{Magnetic field dependence of resonance frequency obtained
in (1) the main mode of ST-FMR measured at $I_{\rm DC} =$ 0 mA
(blue inverted triangle), (2) auto-oscillation (AO) modes of $n =
1$ (red circle), $n = 2$ (square orange), and $n = 3$ (green
triangle) measured at $I_{\rm DC} =$ -7 mA, and (3) linear spin
wave (SW) eigenmodes (red, orange, and green solid lines for $n =
1, 2$, 3 respectively) obtained in the micromagnetic simulations
using Mumax3 with $I_{\rm DC} = 0$ mA and $4 \pi M_{s} =$ 6.5 kG.
The frequency discrepancy $\sim$0.5 GHz between the linear spin
wave modes and the AO mode can be explained by the static
magnetization reduction and the frequency jump near the onset
$I_{\rm th}$ that has been observed as a feature of
self-localization in the nonlinear \textquotedblleft
bullet\textquotedblright modes.} \label{fig:Hvsf0}
\end{figure}

Fig. \ref{fig:Hvsf0} shows the resonance frequency $f_0$ of the
auto-oscillation as a function of the applied magnetic field $H_0$
compared to the linear spin wave eigenmode frequency and the
resonance frequency of ST-FMR at $I_{\rm DC} = 0$ mA. The AO mode
appears below the corresponding linear spin wave mode with the
discrepancy of $\sim$0.5 GHz due to two contributions: One by
$\sim$0.2 GHz is the reduction of static magnetization due to
increased cone angle shown in Fig. \ref{fig:fmr parameters}(b) and
\ref{fig:fmr parameters}(g), and the other by $\sim$0.3 GHz is the
frequency jump of the AO mode from the linear spin wave mode
frequency occurring near the onset $I_{\rm th}$. The latter has
been observed as the signature of the bullet mode in
self-localization \cite{Demidov2012,Demidov2014,Slavin2005}.

\section{Temperature dependence of the intrinsic threshold
current}\label{section:intrinsicIth}

\begin{figure}
\centering
\includegraphics[width=1\columnwidth]{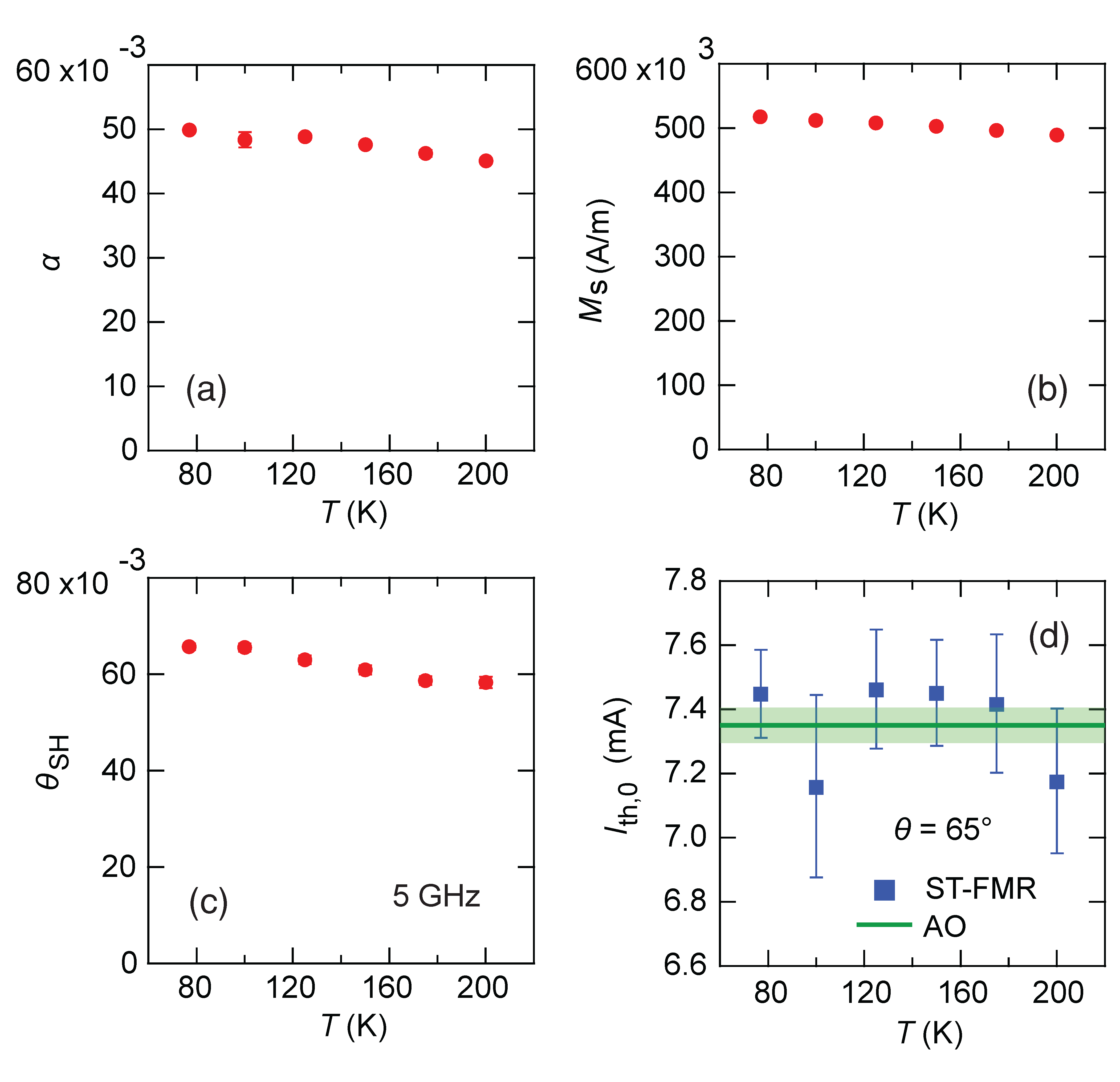}
\caption{Temperature dependence of (a) Gilbert damping constant
$\alpha$, (b) saturation magnetization $M_{\rm S} \left( \simeq
M_{\rm eff} \right)$, (c) spin Hall angle $\theta_{\rm SH}$,
obtained from ST-FMR. (d) Temperature dependence of the intrinsic
threshold current $I_{\rm th,0}$ induced by the spin Hall effect
estimated by Eq. (\ref{eqn:Ith0_T}) with parameters obtained from
the ST-FMR in (a)-(c) in comparison with that estimated from the
AO data in Fig. \ref{fig:nonlinear parameters}(c) of the main
text.}\label{fig:Tdep_Ith0}
\end{figure}

The threshold current $I_{\rm th}$ we obtain from measurement of
the AO power has two contributions as expressed in Eq.
(\ref{eqn:Ith}) of the main text: i) that due to thermal
fluctuation, $\kappa_{\rm th} T$, and ii) an intrinsic threshold
current $I_{\rm th,0}$ arising from the anti-damping torque due to
the applied spin current. Here we estimate the temperature
dependence of the intrinsic threshold current $I_{\rm th,0} \left(
T \right)$ from the ST-FMR data. According to the
Slavin-Tiberkevich theory described in Appendix \ref{section:Q},
the intrinsic $I_{\rm th,0}$ arising from the spin Hall effect is
given by
\begin{equation}
I_{\rm th,0} =  \frac{\alpha \omega}{\sigma} \label{eqn:Ith0}
\end{equation}
The spin-current efficiency $\sigma$ is proportional to the
temperature-dependent $\theta_{\rm SH} / M_{\rm S}$ in our AO
system \cite{Liu2011}.
\begin{equation}
\sigma \left( T \right) \propto  \frac{\theta_{\rm SH} \left( T
\right)}{M_{\rm S} \left( T \right)} \label{eqn:sigma}
\end{equation}
From Eq. (\ref{eqn:Ith0}) and (\ref{eqn:sigma}), the intrinsic
threshold is given as
\begin{equation}
I_{\rm th,0} \left( T \right) = C \frac{\alpha \left( T \right)
M_{\rm S} \left( T \right)}{\theta_{\rm SH} \left( T \right)}
\label{eqn:Ith0_T}
\end{equation}
where $C$ is a temperature independent constant. $\theta_{\rm SH}$
has been found to be temperature independent with a value of 0.068
for 13 - 300 K \cite{Wang2014}, and $\alpha$ for Py film with same
thickness 5 nm is also almost constant in the temperature range of
5 - 300 K regardless of capping materials even though there is a
slight temperature dependence of surface damping around 50 K
\cite{Zhao2016}. Also, $M_{\rm S}$ can be considered to be almost
constant at $T \ll T_{\rm C}$. Therefore, $I_{\rm th,0}$ can be
taken to be almost constant for varying temperature. Based on
this, we express $I_{\rm th} \left( T \right)$ determined from our
AO power in Fig. \ref{fig:nonlinear parameters} as a simple sum of
the temperature independent $I_{\rm th,0}$ and additional thermal
fluctuation contribution linearly decreasing with temperature as
Eq. (\ref{eqn:Ith}) in the main text. We also measured almost
constant $\alpha$, $M_{\rm S}$, and $\theta_{\rm SH}$ from the
ST-FMR in the temperature range of 77 - 200 K as shown in Fig.
\ref{fig:Tdep_Ith0} (a)-(c), and Fig. \ref{fig:Tdep_Ith0}(d) shows
our estimated temperature dependence of intrinsic threshold
$I_{\rm th,0} \left( T \right)$, which is almost constant over the
temperature within our estimation uncertainty.

\appendix

\bibliography{PyPtAO}

\end{document}